\documentclass[onecolumn,amsmath,amssymb,nofootinbib,12pt]{article}
\usepackage{jheppub}

\usepackage{graphicx}
\usepackage{dcolumn}
\usepackage{bm,psfrag}

\usepackage{subfigure}
\usepackage{float}
\usepackage{tensor}

\newcommand{\ben}{\begin{equation}}
\newcommand{\een}{\end{equation}}
\newcommand{\be}{\begin{equation}}
\newcommand{\ee}{\end{equation}}
\newcommand{\bea}{\begin{eqnarray}}
\newcommand{\eea}{\end{eqnarray}}
\newcommand{\ba}{\begin{eqnarray}}
\newcommand{\ea}{\end{eqnarray}}

\newcommand{\beq}{\begin{equation}}
\newcommand{\eeq}{\end{equation}}
\newcommand{\beqa}{\begin{eqnarray}}
\newcommand{\eeqa}{\end{eqnarray}}
\newcommand{\beqar}{\begin{eqnarray*}}
\newcommand{\eeqar}{\end{eqnarray*}}

\newcommand{\reef}[1]{(\ref{#1})}

\newcommand{\eg}{{\it e.g.,}\ }
\newcommand{\ie}{{\it i.e.,}\ }
\newcommand{\comment}[1]{{\bf [[[#1]]]}}

\newcommand{\cO}{{\cal O}}

\newcommand{\labell}[1]{\label{#1}} 

\newcommand{\lp}{\ell_{\mt P}}

\def\t6 {T_\mt{D6}}

\newcommand{\mt}[1]{\textrm{\tiny #1}}

\newcommand{\ve}{\varepsilon}

\def\cale         {{\cal E}}

\def\calo         {{\cal O}}

\def\ee           {{\rm e}}

\def\sqr#1#2{{\vcenter{\vbox{\hrule height.#2pt
 \hbox{\vrule width.#2pt height#1pt \kern#1pt
 \vrule width.#2pt}\hrule height.#2pt}}}}

\def\ee{\cale}

\def\hg{\hat{g}}

\def\aa1{\phi}
\def\cc1{\psi}

\def\vev#1{\langle #1 \rangle}

\def\nnn{\nonumber}

\def\comment#1{{\bf [[#1]]}}

\newcommand{\tlam}{\tilde{\Lambda}}

\begin{document}

\preprint{arXiv:1601.nnnnn [hep-th]}

\title{Comments on Jacobson's ``Entanglement equilibrium and the Einstein equation"}

\author{Horacio Casini$^1$,}
\author{Dami\'an A. Galante$^{2,3}$,}
\author{Robert C. Myers$^{3}$}
\affiliation{$^1$\, Centro At\'omico Bariloche and CONICET,\\ 
\vphantom{k}\ \ S.C. de Bariloche, Rio Negro, R8402AGP, Argentina}
\affiliation{$^2$\,Department of Applied Mathematics, University of Western Ontario,\\ 
\vphantom{k}\ \ London, ON N6A 5B7, Canada}
\affiliation{$^3$\,Perimeter Institute for Theoretical Physics, Waterloo, ON N2L 2Y5, Canada}

\emailAdd{casini@cab.cnea.gov.ar}
\emailAdd{dgalante@perimeterinstitute.ca}
\emailAdd{rmyers@perimeterinstitute.ca}

\date{\today}

\abstract{Using holographic calculations, we examine a key assumption made in Jacobson's recent argument for deriving Einstein's equations from vacuum entanglement entropy. Our results involving relevant operators with low conformal dimensions seem to conflict with Jacobson's assumption. However, we discuss ways to circumvent this problem.
}

\maketitle

\section{Introduction and Summary of Results}
\label{intro}
Entanglement is now recognized to play an important role in the emergence of space (and space-time) in quantum gravity with calculations from a variety of different approaches  \cite{area,ted0,mvr,arch,holes,cool}. Of course, one of the most natural frameworks where this connection can be investigated is the AdS/CFT correspondence \cite{revue}. In particular, the elegant prescription for holographic entanglement entropy \cite{rt1} reveals a deep connection between entanglement entropy and spacetime geometry \cite{mvr}. However, this connection was further extended to the dynamics of the spacetime in \cite{wow} which related the first law of entanglement in the vacuum of the boundary CFT to the Einstein equations linearized around the AdS vacuum in the bulk. Of course, the latter is reminiscent of Jacobson's derivation of Einstein's equations from thermodynamic arguments involving Rindler horizons in \cite{ted0}. 

More recently, Jacobson \cite{ted} proposed an intriguing argument in which the full nonlinear Einstein equations arise from the postulate that the vacuum entanglement entropy for small balls is extremal. This argument  makes a precise and profound connection between entanglement and gravity. The aim of this paper is to examine a certain key assumption that was needed for this argument to hold. In particular, it was assumed that the variation of the entanglement entropy for quantum fields in small casual diamonds takes a specific form. In the following, we test this assumption using holographic calculations but first let us briefly review Jacobson's argument:

Jacobson \cite{ted,ted2} begins by considering the entanglement entropy of a small (space-like) spherical region $\cal R$ in the vacuum and makes small deformations of the geometry $\delta g_{ab}$ and of the state of the matter fields $\delta |\psi\rangle$. With the assumption that the vacuum entanglement entropy is extremal, the variation of entanglement entropy then vanishes (to leading order) through the cancellation of two contributions: 
\begin{eqnarray}
\delta S_{total} = \delta S_\mt{UV} + \delta S_\mt{IR} = 0\,. \labell{eqdeltaS}
\end{eqnarray}
Here, $\delta S_\mt{UV}$ is a universal UV contribution, arising from the change in geometry. Quantum gravity is expected to render this UV contribution finite and produce a result proportional to the change in the boundary area, \ie $\delta S_\mt{UV}=\delta A/(4G)$. The IR contribution $\delta S_\mt{IR}$ arises from the change in the state of the matter fields. The crucial point for the remaining discussion is the precise form of this IR variation.

Quite generally, the first law of entanglement \cite{relative} allows the leading contribution to the latter to be written as
\begin{eqnarray}
\delta S_\mt{IR} =  \delta\vev{\cal H}\,,
\labell{firstlaw}
\end{eqnarray}
where $\cal H$ is the modular Hamiltonian for the density matrix $\rho$ produced when the global state is reduced to a given region of interest, \ie ${\cal H}=-\log\rho$.  However, this result is not particularly useful except in special cases. One such special case arises when the vacuum state of a CFT in flat space is reduced to a spherical region. In this case, the modular Hamiltonian is given by the integral of the energy density $T_{00}$ with a simple profile across the spherical region \cite{CHM} --- see eq.~\reef{modu} below. 

In his argument, Jacobson chooses the radius $R$ of the sphere to be much smaller than any (length) scale in the geometry or in the quantum field theory, but still much larger than Planck scale $\lp$ --- see section \ref{discuss} for further discussion. Hence in this small sphere limit, the energy density is essentially constant throughout this region and the first law \reef{firstlaw} yields 
\begin{eqnarray}
\delta S_\mt{IR,\,CFT} = 2\pi\,\frac{\Omega_{d-2}\, R^d}{d^2-1}\ \delta\vev{T_{00}} \,,
\labell{IRS}
\end{eqnarray}
where $\delta\vev{T_{00}}$ is the change in the energy density in comparison to the vacuum state.
Further, $d$ is the spacetime dimension, and $\Omega_{d-2}=2\pi^{\frac{d-1}2}/\Gamma\left(\frac{d-1}2\right)$ is the volume of a $(d$--2)-dimensional unit sphere. Next, evaluating the expression for the change in the boundary area at fixed volume in some maximally symmetric reference geometry, one finds a result involving the time-time component of the Einstein tensor. Combining this result and the above expression \reef{IRS} for $\delta S_\mt{IR}$ of spheres in all reference frames and positions, one can then derive the Einstein equations (with a cosmological constant) for a CFT coupled to gravity \cite{ted2},
\begin{eqnarray}
G_{ab} + \Lambda\, g_{ab} = 8 \pi G \  \delta\vev{T_{ab}} \,. \labell{Einstein}
\end{eqnarray}

One would like to extend the above discussion to the case of a nonconformal quantum field theory. In this general case, Jacobson \cite{ted,ted2} makes the assumption that eq.~\reef{IRS} extends to the following form:
\begin{eqnarray}
\delta S_\mt{IR} = 2\pi\, \frac{\Omega_{d-2} R^d}{d^2-1}\, \Big(\, \delta\vev{T_{00}} - \delta\vev{X}\,  \Big) \,,
\labell{tedsproposal}
\end{eqnarray}
where $X$ is some scalar operator in the QFT. For example, an initial suggestion was that $X=-T^a{}_a/d$ in which case, the expression on the right-hand side of this equation is proportional to (the time-time component of) the traceless part of the stress tensor \cite{ted}. In any event, with this assumption \reef{tedsproposal}, it is straightforward to generalize the arguments above and derive eq.~\reef{Einstein} for a general quantum field theory coupled to gravity.
Although the spacetime geometry is considered dynamical here, we emphasize that eq.~\reef{tedsproposal} is a conjecture for quantum field theories in a fixed flat spacetime.\footnote{The latter follows since Jacobson considers a small sphere limit where $R$ much smaller than any (length) scale in the geometry or in the QFT, but still $R\gg\lp$.}

In this paper, we analyze the validity of this key assumption \reef{tedsproposal} within the context of the AdS/CFT correspondence. In particular, we consider a {\it holographic} CFT perturbed by a relevant operator $\calo_\Delta$ of scaling dimension $\Delta$. That is, the action of the boundary theory becomes
\begin{eqnarray}
I=I_{CFT} + \int d^d x \ \lambda \ \calo_{\Delta}(x)\,.
\labell{perturb}
\end{eqnarray}
with some (dimensionful) coupling $\lambda$.
The dual gravitational problem corresponds to solving the Einstein equations coupled to a massive scalar field (and a negative cosmological constant) in $d+1$ dimensions. Further, we can compute the entanglement entropy by using the holographic prescription of \cite{rt1}. Since we are only considering small spherical regions, the holographic entanglement entropy is only probing the asymptotic geometry of the bulk spacetime and we can proceed by only considering the asymptotic behaviour of the bulk fields. The metric perturbation and bulk scalar become vanishingly small as we approach the asymptotic boundary and hence we calculate perturbatively to leading order in the amplitude in the scalar field. In terms of the boundary theory, $R^d\vev{T_{ab}}$, $R^\Delta \vev{\calo_{\Delta}}$ and $R^{d-\Delta}\lambda$ are all small and we work to leading order in these (dimensionless) quantities. The details of the calculations will be described  below in the next section, but let us present here the main results coming from this analysis.
%

For relevant operators with $\frac{d}2<\Delta<d$, we find
\begin{eqnarray}
\delta S_\mt{IR}  & = & 2\pi\,\frac{\Omega_{d-2} R^d}{d^2-1} \left( \delta\vev{T_{00}} - \frac{1}{2 \Delta - d}\, \delta\vev{T^a{}_a} 
\right)  \nnn \\
& &\qquad -   \frac{2^{d-2}d(d+1)\Delta\ \Gamma\!\left(\frac{d-1}{2}\right) \Gamma\!\left(\Delta+1-\frac{d}{2} \right)}{(d-\Delta)^2\ \Gamma\!\left(\Delta +\frac{3}{2}\right)}\ \frac{R^{2\Delta}\,\delta\vev{\calo_\Delta}^2}{C_T} +\cdots\, , \labell{formula}
%
\end{eqnarray}
where $C_T$ is the central charge appearing in the vacuum correlator of the stress tensor  --- see eq.~\reef{emt2p} below. 
In the deformed boundary theory, various expectation values may be nonvanishing in the vacuum state and hence our notation above indicates, \eg $\delta\vev{T_{00}}\equiv\vev{T_{00}}-\vev{T_{00}}_{\rm vacuum}$.\footnote{For clarity, let us add that our notation is that $\delta\,\vev{\calo_\Delta}^{2}=\vev{\calo_\Delta}^{2}-\vev{\calo_\Delta}^{2}_{\rm vacuum}$ above and throughout the paper.} Now the first line in the above expression matches the desired form given in eq.~(\ref{tedsproposal}). However, we note that it is not the traceless part of the stress tensor that appears here. We have also included an extra contribution in the second line of eq.~(\ref{formula}), which is proportional to $\delta\,\vev{\calo_\Delta}^{2}$ and, as dictated by dimensional analysis, this term is accompanied by a factor $R^{2\Delta}$. Now the expectation values, $\delta\vev{T_{ab}}$ and $\delta\,\vev{\calo_\Delta}^2$ are determined by infrared scales that are independent of $R$ and hence because $d<2\Delta$ as we take a limit $R\to0$, the contribution in the second line becomes negligible compared to the  contributions involving $\delta\vev{T_{ab}}$. Hence this result \reef{formula} agrees with the form of $\delta S_\mt{IR}$ required in the derivation of Einstein's equations.

Note that the coefficient of the $\delta\vev{T^a{}_a}$ in eq.~\reef{formula} is singular in the limit $\Delta\to d/2$. Hence the holographic calculations must be redone for the particular case of $\Delta=d/2$ --- see section \ref{deltaover2} --- with the result
\begin{eqnarray}
\delta S_\mt{IR} &=& \frac{\Omega_{d-2} R^d}{d^2-1} \left( \delta \vev{T_{00}} + \delta \vev{T^a{}_a} \left( \frac{d+2}{d(d+1)} - \frac{1}{2} H_{\frac{d-1}{2}}+\log (\mu R) \right)\right)\nonumber\\
&&\qquad\qquad\qquad\qquad  -\frac{2^{d-1}\,d^2}{d-1}\, \frac{R^d\,\delta \vev{\calo_\Delta}^2}{C_T}+\cdots\,, \labell{formula2}
\end{eqnarray}
where $H_{\frac{d-1}{2}}$ is the harmonic number defined by $H_n =\int_0^1 dx \frac{1-x^n}{1-x}$. Also note that we have a new renormalization scale $\mu$ appearing in  the logarithmic term. In this case, the dimension of $\delta\vev{\calo_\Delta}^{2}$ matches that of the stress-energy tensor and so at this order, all of the contributions appear with the same overall factor of $R^d$. Hence eq.~\reef{formula2} almost has the desired form \reef{tedsproposal} except for the appearance of an extra logarithmic dependence on $R$ in the term proportional to $\delta \vev{T^a{}_a}$.

Now the unitarity bound in CFTs also allows for scalar operators with conformal dimension $(d-2)/2 < \Delta < d/2$. However, this regime requires the so-called alternative quantization of the holographic theory \cite{alter}, which will subsequently involve an alternative holographic renormalization procedure.  Interestingly, this procedure yields the same result as for $\Delta>d/2$. That is, $\delta S_\mt{IR}$ is still given by eq.~(\ref{formula}) in this new regime. However, with $\Delta<d/2$, the term proportional to $R^{2\Delta}$ becomes the leading contribution in the small $R$ expansion. Hence our holographic results for operators in this regime present a challenge for Jacobson's argument. However, let us emphasize that the contribution proportional to $R^{2\Delta}\delta\vev{\cO_\Delta}^2$ in eq.~\reef{formula} 
goes beyond the first law variation \reef{firstlaw} which was central to Jacobson's reasoning. In section \ref{discuss}, we discuss this point and several ways in which, in principle, this extra contribution can be incorporated into Jacobson's argument so that one could still derive Einstein's equations from an principle of maximal vacuum entanglement.

The remainder of the paper is organized as follows: We begin in section \ref{main} by describing the details of the holographic model and of our holographic entanglement entropy calculations. In subsection \ref{general}, we compute $\delta S_\mt{IR}$ for operators with conformal dimension in the regime $d/2<\Delta<d$, while subsections \ref{alternate} and \ref{deltaover2} describe the computations for the regime $(d-2)/2<\Delta<d/2$ and the specific case $\Delta=d/2$, respectively. In section \ref{discuss}, we conclude with a brief discussion of the implications of our results for the proposal in \cite{ted,ted2}. We have three appendices to discuss certain technical details. Appendix \ref{alt_holo_ren} describes the details of holographic renormalization in the context of the alternate quantization. Appendix \ref{slider} presents a short calculation of
the contribution in the shift in the entanglement entropy proportional to $\vev{T_{00}}^2$ for a thermal state in a $d=2$ CFT. 
Appendix \ref{higher} extends the calculation of $\delta S_\mt{IR}$ to include the next-to-leading order contributions in the coupling $\lambda$.

\section{Main Results}
\label{main}
In this section, we provide the details of the calculations that produced the results given in eqs.~\reef{formula} and \reef{formula2} above. We start by setting up our  holographic model in subsection \ref{general}. The results in that section are valid for general values of the scaling dimension $\Delta$. However, certain values require extra consideration; this is presented in subsections \ref{alternate} and \ref{deltaover2} --- see also appendix \ref{alt_holo_ren}. In appendix \ref{higher}, we also present calculations to next order in the perturbation parameter and compute subleading contributions in the variation of the entanglement entropy.

\subsection{Holographic Framework}
\label{general}
Our holographic model is comprised of the following action,
\begin{eqnarray}
I_{bulk}= \frac{1}{2  \lp^{d-1}} \int d^{d+1}x  \sqrt{-G} \left[ R - \frac{1}{2} (\nabla \Phi)^2 - V(\Phi) \right],
\labell{ibulk}
\end{eqnarray}
where
\begin{eqnarray}
V(\Phi) = - \frac{d (d-1)}{L^2} + \frac{1}{2} m^2 \Phi^2 + \frac{\kappa}{6 L^2} \Phi^3 + O(\Phi^4)\,.\labell{potential}
\end{eqnarray}
Here for completeness, we included a cubic term in the potential but this term will be neglected in all of our calculations in the main text and it will only play a role in Appendix \ref{higher}. Of course, if $\Phi=0$, the metric solution for the vacuum will be pure $AdS_{d+1}$ with $L$ being the curvature scale. However, in general, our calculations involve examining the Einstein and scalar field equations together and finding solutions where the scalar has a nontrivial profile reflecting the presence of the relevant perturbation \reef{perturb} in the boundary theory. As noted previously, we will only be examining the asymptotic region near the boundary of the bulk spacetime where the metric perturbation and bulk scalar become vanishingly small. This allows us to construct the solutions perturbatively in the amplitude in the scalar field. 

Near the asymptotic boundary, it is useful to introduce Fefferman-Graham coordinates for the metric,
\begin{eqnarray}
ds^2 = \frac{L^2}{z^2} \left(dz^2 + g_{ab}(z,x)\, dx^a dx^b \right)\,,
\labell{feffer}
\end{eqnarray}
where $z$ is the holographic coordinate with $z=0$ corresponding to the boundary. Near the boundary, we can expand $g_{ab}(z,x)$ as
\begin{eqnarray}
g_{ab} (z) =\ \eta_{ab} + \sum_k g_{ab}^{(k)}(x)\, z^k \,,\labell{fgump}
\end{eqnarray}
where the first term (\ie $\eta_{ab}$) is chosen for a flat boundary metric. The standard AdS/CFT dictionary relates the expectation value of the boundary stress tensor with the $z^d$ coefficient in this expansion, as $\vev{T_{ab}} = \frac{d}{2 \, \lp^{d-1} \, L} \, g_{ab}^{(d)}$. However, for the nonconformal case, we reconsider this expression in section \ref{holren}. Similarly, the scalar field has two independent asymptotic solutions,
\begin{eqnarray}
\Phi(z) \sim \phi_0(x)\, z^{d-\Delta} + \phi_1(x)\, z^{\Delta}\,, \labell{such}
\end{eqnarray}
where, as the notation indicates, the exponent $\Delta$ is the conformal dimension of the dual operator and is given by
\begin{eqnarray}
\Delta = \frac{d}{2} + \sqrt{\frac{d^2}{4} + m^2 L^2}\,.
\labell{mass}
\end{eqnarray}
Now, the usual holographic dictionary relates the first coefficient $\phi_0$ to the coupling $\lambda$, and $\phi_1$ to the expectation value of the relevant operator $\calo_\Delta$ --- the explicit relations will be given below in section \ref{holren}. Implicitly with eq.~\reef{mass} and throughout this subsection, we are assuming the relevant operators have $d/2 < \Delta < d$. As commented before, the cases of $(d-2)/2 < \Delta < d/2$ and $\Delta=d/2$ will be treated separately in subsequent subsections.

In the above discussion, the ansatz has been fairly general so that it could describe expectation values which vary across the boundary spacetime, \eg $\vev{\cO_\Delta}(x)$. However, in the problem of interest, we want to probe the boundary theory by examining the entanglement entropy of spheres that are much smaller than the scale of any such variations. Hence for simplicity, we will assume that our holographic background is invariant under translations in both space and time in the following. Hence the metric perturbations $g_{ab}^{(k)}$ in eq.~\reef{fgump} and the coefficients $\phi_{0,1}$ in the scalar \reef{such} are constants, and our metric ansatz in eq.~\reef{feffer} is simplified with $g_{ab}(z,x)$ reduced to $g_{ab}(z)$. We also impose the boundary condition that $g_{ab}(z)\to\eta_{ab}$ asymptotically to recover the flat boundary metric in eq.~\reef{fgump}. Note, however, that we are not otherwise restricting the metric perturbations, which will allow us to consider states which are anisotropic and stationary (but not static). 

With these choices, the Einstein equations become, \eg \cite{skinny,HMS}:
%
%
\begin{eqnarray}
0&=&g''_{ab}-\frac{d-1}zg'_{ab}-g^{cd}g'_{ca}\,g'_{db}+
\frac12g^{cd}g'_{cd}\,g'_{ab}
-\frac{1}z g^{cd}g'_{cd}\, g_{ab}+\frac{\,g_{ab}}{(d-1)z^2}\left(m^2L^2\Phi^2+\frac{\kappa}{3}
\Phi^3\right)
 \nonumber\\
0&=&g^{ab}g''_{ab}-\frac1z\,g^{ab}g'_{ab}-\frac{1}2 g^{ab}g'_{bc}g^{cd}g'_{da}+\Phi^{\prime\,2}+\frac{1}{(d-1)z^2}\left(m^2L^2\Phi^2+\frac{\kappa}{3}
\Phi^3\right)
\labell{einstein1}
\end{eqnarray}
Similarly, the scalar field equation following from eqs.~\reef{ibulk} and \reef{potential} becomes
\begin{eqnarray}
\Phi ''-\frac{(d-1) \Phi '}{z}
+\frac{g^{ab}\,g_{ab}'}2\,\Phi'-\frac{(m L)^2 \Phi }{z^2}-\frac{\kappa  \Phi ^2}{2 z^2} =0.
\labell{KGeq2}
\end{eqnarray}
Moreover, we note that these equations are redundant. For example, the second equation in eq.~\reef{einstein1} will automatically be solved if the first set of equations, as well as the scalar field equation \reef{KGeq2}, are solved. However, this equation may still be used as a consistency check for the solutions.

As commented above, we will construct the solutions perturbatively in the amplitude in the scalar field. Hence we introduce a small expansion parameter $\ve$, so that the scalar is written as
\begin{eqnarray}
\Phi(z) = \ve \, \phi_0 \, z^{d-\Delta} + \ve \, \phi_1 \, z^{\Delta} + O(\ve^2) \labell{such2}
\end{eqnarray}
as $z$ approaches zero --- we re-iterate that $\phi_0$ and $\phi_1$ are now simply constants. Solving eq.~(\ref{KGeq2}) to first order in $\ve$ yields the usual solution for $\Delta$ given in eq. (\ref{mass}).
Next, we solve eq.~(\ref{einstein1}) to second order in $\ve$. Here, we formulate the most general ansatz for the metric functions $g_{ab}(z)$ which approach $\eta_{ab}$ as $z\to0$,
\begin{eqnarray}
g_{ab} = \eta_{ab} + \ve^2 \left( m_{ab} \, z^d + a_{ab} \,  \phi_0^2 \, z^{2\Delta} + b_{ab} \, \phi_1^2 \, z^{2(d-\Delta)} + c_{ab} \, \phi_0 \, \phi_1 \, z^d \right) + O(\ve^3) \,,\labell{juk}
\end{eqnarray}
where $m_{ab}$, $a_{ab}$, $b_{ab}$ and $c_{ab}$ are all matrices with constant coefficients. The role of the $m_{ab}$ terms will be to introduce additional contributions to the stress-energy tensor which are independent of the conformal perturbation. Implicitly, we have also set $L=1$ here. Substituting this
ansatz \reef{juk} into the Einstein equations (\ref{einstein1}), we find the solution 
\begin{eqnarray}
a_{ab} & = & b_{ab} = - \frac{\eta_{ab} }{4 (d-1)}\,, \labell{brown}\\
c_{ab} & = & -\frac{2\Delta (d-\Delta )}{d^2(d-1)} \,\eta_{ab} \,,\qquad
m^a{}_a  =  0 \,.\nonumber
\end{eqnarray}
%
%
The trace in the last term is made with the boundary metric, \ie $m^a{}_a=\eta^{ab}\,m_{ab}$. Here, we must comment that there is some ambiguity in the metric ansatz \reef{juk} because both $m_{ab}$ and $c_{ab}$ appear at order $z^d$. In particular, if we shift these two matrices by $\delta m_{ab}=-\delta c_{ab}\,\phi_0\phi_1$ and $\delta c_{ab}$, respectively, the metric is left invariant at this order in the $\ve$ expansion. In fact, the Einstein equations \reef{einstein1} only fix the trace of the $z^d$ contribution in eq.~\reef{juk} and hence, in writing the second line of eq.~\reef{brown}, we are making a convenient choice for these matrices which simplifies our results in the following. 

\subsubsection{Holographic Renormalization}
\label{holren}

The ultimate aim is to express the variation of the entanglement entropy for a spherical region in terms of field theoretic quantities. In this section, we apply the usual holographic renormalization to evaluate various expectation values in the field theory in terms of dual parameters in the gravitational solution. 

First, we must evaluate the on-shell bulk action but to regulate the result, we introduce a cut-off surface at $z=z_\epsilon$ near the boundary. Then, we add a counterterm action that cancels the divergences \cite{count}. Hence the total gravitational action becomes
\begin{eqnarray}
I_{reg} = I_{bulk} + I_{GHBY} + I_{ct},
\end{eqnarray}
where $I_{bulk}$ is just the action in eq. (\ref{ibulk}) and the two boundary contributions are
\begin{eqnarray}
I_{GHBY} & = & - \frac{1}{\lp^{d-1}} \int d^d x \sqrt{-\gamma}\, K \rvert_{z=z_\epsilon} \, ,\labell{boundary} \\
I_{ct} & = & - \frac{1}{2 \lp^{d-1}} \int d^d x \left. \sqrt{-\gamma} \left( 2(d-1) + \frac{d-\Delta}{2} \Phi^2  \right) \right\rvert_{z=z_\epsilon} \, .\nonumber
\end{eqnarray}
In these expressions, $\gamma$ is the (determinant of the) induced metric on the cut-off surface. In general, the counterterm action will contain extra terms involving boundary curvatures and derivatives of the scalar field, \eg see \cite{count,anton}. However, eq.~\reef{boundary} is sufficient for the present purposes. 
Now we are interested in finding the expectation value of the stress-energy tensor and the operator $\calo_\Delta$, so we take functional derivatives of the action, \ie
\begin{eqnarray}
\vev{T^{ab}} & = & \lim_{z_\epsilon \to 0} \frac{2}{ \sqrt{-g^{(0)}}} \frac{\delta I_{reg}}{\delta g_{ab}^{(0)}} \, ,\labell{expect}\\
\vev{\calo_\Delta} & = & \lim_{z_\epsilon \to 0} \frac{1}{\sqrt{-g^{(0)}}} \frac{\delta I_{reg}}{\delta \lambda} \, ,\nonumber
\end{eqnarray}
where $\lambda=\ve\,\phi_0$. The limit $z_\epsilon \to 0$ yields the (finite) renormalized expectation values.

We performed the above calculations following appendix C of \cite{anton} and found
\begin{eqnarray}
\vev{T_{ab}}&=&\frac{\ve^2}{2 \lp^{d-1}} \left( d \, m_{ab}+ \eta_{ab} \frac{(d-\Delta)(2\Delta-d)}{d} \phi_0 \, \phi_1 \right)  + O(\ve^3) \, ,\labell{stressed}\\
\vev{T^a{}_a} & = & \frac{\ve^2}{2 \lp^{d-1}} (d-\Delta) (2\Delta -d) \phi_0 \, \phi_1 + O(\ve^3) \, , \labell{tracet} \\
\vev{\calo_\Delta} & = & \frac{\ve}{2 \lp^{d-1}} ( 2\Delta-d) \phi_1 + O(\ve^2) \, . \labell{vevO}
\end{eqnarray}
For convenience, we have also included the trace of the stress tensor above --- as before, the trace is performed with the flat boundary metric, \ie $\vev{T^a{}_a} = \eta^{ab}\,\vev{T_{ab}}$.

Finally, it is useful to consider the central charge $C_T$ which appears in the two-point function of the stress tensor  \cite{Osborn0,Erdmenger0}, 
 \beq
  \langle  T_{ab}(x)\,  T_{cd}(0)  \rangle = {C_T \over x^{2d}}\, 
\mathcal{I}_{ab,cd}(x)~,
  \labell{emt2p}
 \eeq
where the structure of $\mathcal{I}_{\mu\nu,\alpha\beta}(x)$ is completely fixed by conformal invariance. This expression applies in the vacuum state on $R^d$ for any general CFT. Holographic calculations of this correlator then yield, \eg \cite{twist}
\beq
C_T = \frac{2^{d-1}d(d +1)}{\pi\,\Omega_{d-2} } \ \frac{1}{\lp^{d-1}} \,,
\labell{relations}
\eeq 
for the boundary CFT dual to eq.~\reef{ibulk}. Recall that we have set the AdS scale $L=1$ and hence the last factor implicitly corresponds to $(L/\lp)^{d-1}$ --- in fact, the same factor appears in each of the expectation values above in eqs.~(\ref{stressed}--\ref{vevO}).

\subsubsection{Entanglement Entropy Calculation}
\label{eec}

We want to compute the holographic entanglement entropy for a spherical boundary region $\cal R$ of radius $R$. According to the Ryu-Takayanagi prescription \cite{rt1}, this is given by
\begin{eqnarray}
S = \frac{2 \pi}{\lp^{d-1}}\ \underset{v\sim {\cal R}}{\text{ext}} A(v)\,,
\end{eqnarray}
where $A(v)$ is the area of the ($d$--1)-dimensional bulk surface $v$ and we extremize over all such surfaces which are homologous to $\cal R$.

In the AdS vacuum, the extremal surface $v$ for a spherical entangling surface is well known \cite{rt1}. In this case, the bulk metric \reef{feffer} simplifies with $g_{ab}(z)=\eta_{ab}$ and one can take advantage of the spherical symmetry to consider a bulk profile $z(r)$ where $r^2 = \sum (x^i)^2$. The minimal surface is just the hemisphere: $z_0^2 = R^2-r^2$. In the present case, we should extremize the area functional with a general profile $z(x)$ in the bulk geometry defined by our general ansatz for $g_{ab}(z)$. However, we are only working to leading order in a perturbative expansion around the vacuum AdS geometry. In this case, one can show that any change in the position of the surface will not contribute to the first order correction in the value of $A(v)$, \eg see \cite{relative}.  Hence it is sufficient to compute $A(v)$ with the vacuum profile $z_0(r)$ but with the perturbed metric functions $g_{ab}(z)$.

In this case, the induced metric on the bulk surface becomes\footnote{Here and in the following, the implicit sums over Latin indices in the middle of the alphabet only run over the spatial directions, \ie $i,j = 1,2,\cdots,d-1$.}
\begin{eqnarray}
h_{ij}dx^i dx^j = \frac{1}{z_0^2}  \left ( g_{ij}(z_0) + z_0^{\prime\,2} \mu_i\mu_j  \right)  dx^i \, dx^j  \,,
\end{eqnarray}
where, as described above, $z_0^2=R^2-r^2$ and $z_0'=dz_0/dr$. Further $\mu_i$ are direction cosines for the spatial coordinates, \ie $x^i= r\mu_i$ with the normalization $\sum \mu_i^2 =1$, \eg see \cite{spike}. Eq.~\reef{juk} gives the background metric but for simplicity, we combine the various metric perturbations as 
\beq
 g_{ij}(z_0) =\delta_{ij} +\ve^2\, \hg_{ij}(z_0) + \cO(\ve^4)\,.
\labell{ohhh}
\eeq
Then to leading order, the area functional becomes 
\begin{eqnarray}
A(v) &=& \int_{r\le R}\!\!\!\! d^{d-1}x \sqrt{{\rm det}h_{ij}}=\int_{r\le R}\!\! \frac{r^{d-2}dr\,d\Omega_{d-2}}{z_0^{d-1}}\left[\sqrt{1+z_0^{\prime\,2}}
\vphantom{\frac{\ve^2}{\sqrt{1+z_0^{\prime\,2}}}}
\right.\labell{area8}\\
&&\qquad\quad\left.+
\frac{\ve^2}{\sqrt{1+z_0^{\prime\,2}}} \left(\frac12\sum_i \hg_{ii}(z_0)\,(1+z_0^{\prime\,2}(1-\mu_i^2))+\sum_{i<j}(-)^{i+j}\hg_{ij}(z_0)\,z_0^{\prime\,2}\mu_i\mu_j\right)\right]\,.
\nonumber
\end{eqnarray}
It is straightforward to confirm that $\int d\Omega_{d-2} \mu_i \mu_j=\delta_{ij}\,\Omega_{d-2}/(d-1)$ where as above, $\Omega_{d-2}\equiv\int d\Omega_{d-2}= 2\pi^{\frac{d-1}2}/\Gamma\left(\frac{d-1}2\right)$. Further, it is convenient to replace the radial integral with an integration over the bulk coordinate $z$, which then yields
\begin{eqnarray}
A(v) &=& \Omega_{d-2} \int_0^R \frac{dz\,r_0^{d-2}}{z^{d-1}}\left[ \frac{R}{r_0} 
+
\frac{\ve^2\,r_0}{2\,R}
 \sum_i \hg_{ii}(z)\left(\frac{z^2}{r_0^2}+\frac{d-2}{d-1}\right)\right]
\labell{area86}
\end{eqnarray}
where $r_0^2=R^2-z^2$. Note that implicitly both eqs.~\reef{area8} and \reef{area86} require a UV regulator because various terms in the radial integral diverge as $z\to0$.

We note that gravity is absent in the boundary theory and so our holographic entanglement entropy calculations are only evaluating the variation due to matter fields in a fixed background geometry, \ie we are calculating $\delta S_\mt{IR}$ in eq.~\reef{eqdeltaS}.  In particular, we wish to evaluate $\delta S= S_1- S_0$, where $S_1$ corresponds to the holographic entanglement entropy with the full perturbed metric $g_{ab}(z)$ as in eq.~(\ref{feffer}) and $S_0$ corresponds to that when  the expectation values (\ref{stressed}--\ref{vevO}) have some `vacuum' values. In the latter case, the metric is given by 
\begin{eqnarray}
g_{ab}^{(vac)} = g_{ab}(\phi_1 \to \phi_1^{vac}, m_{ab} \to m_{ab}^{vac} )\,.
\end{eqnarray}
Note that $\delta S$ is finite as all UV divergences in the entanglement entropy correspond to the vacuum divergences and they are cancelled in the difference $S_1-S_0$. Therefore we may ignore the UV cut-off for the radial integral mentioned below eq.~\reef{area86}. Substituting the expressions for $\hat{g}_{ij}(z)$ in eq.~\reef{juk} into eq.~\reef{area86}, as well as $r_0^2=R^2-z^2$, the difference leaves the following expression at order $\ve^2$
\begin{eqnarray}
\delta S & = &  \frac{2 \pi}{\lp^{d-1}} \Omega_{d-2} \ve^2 \int_0^R dz \left[ \frac{ z \left(R^2-z^2\right)^{\frac{d-3}{2}} \left((d-2) R^2+z^2\right)} {2 (d-1)  R}  \right. \labell{minarea} \\
& &\qquad\qquad \times \left(\delta^{ij} \delta m _{ij}-\frac{2\Delta(d-\Delta)}{d^2} \,   \phi _0 \, \delta\phi _1  - \frac14 \, \delta(\phi _1^2) \, z^{d-2\Delta}\right) \Bigg] \nnn \,,
\end{eqnarray}
where as above, we have introduced the notation $\delta X \equiv X - X^{vac}$.
Finally after making the $z$ integration, we obtain
\begin{eqnarray}
\delta S & = & \frac{2 \pi}{\lp^{d-1}} \, \Omega_{d-2} \, \ve^2 \, R^d \frac{d}{2(d^2-1)} \left( \delta m_{00} - \frac{2 \Delta  (d-\Delta)}{d^2}\phi_0 \, \delta\phi_1  \right)  \labell{deltaS} \\
& &\quad - \frac{2 \pi}{\lp^{d-1}} \, \Omega_{d-2} \, \ve^2 \frac{\Delta}{16} \frac{\Gamma \left(\frac{d-1}{2}\right) \Gamma \left(\Delta-\frac{d}{2}+1\right) }{\Gamma \left(\Delta +\frac{3}{2}\right)} \delta(\phi _1^2) \, R^{2 \Delta} + O(\ve^3) \, , \nnn 
\end{eqnarray}
where we have also used the tracelessness of $m_{ab}$ in eq.~\reef{brown} to replace $\delta^{ij} \delta m_{ij} = \delta m_{00}$.

Finally, it is easy to express eq.~(\ref{deltaS}) in terms of field theory quantities using eqs. (\ref{stressed}--\ref{relations}). Hence, we can see  to order $\ve^2$, \begin{eqnarray}
\delta S  & = & 2 \pi \frac{\Omega_{d-2} R^d}{d^2-1} \left( \delta\vev{T_{00}} - \frac{1}{2 \Delta - d} \delta\vev{T^a{}_a}  \right) \labell{jump} \\
& &\qquad -   \frac{2^{d-2}d(d+1)\Delta\ \Gamma\!\left(\frac{d-1}{2}\right) \Gamma\!\left(\Delta+1-\frac{d}{2} \right)}{(d-\Delta)^2\ \Gamma\!\left(\Delta +\frac{3}{2}\right)}\ \frac{R^{2\Delta}\,\delta\vev{\calo_\Delta}^2}{C_T} \, .\nonumber
%
\end{eqnarray}
As noted above, the calculations described in section \ref{holren} are valid for operators with $d/2 < \Delta < d$. Hence our translation of eq.~\reef{deltaS} to eq.~\reef{jump} should only be accepted as valid for this particular range of conformal dimensions. In this case, the contribution proportional to $\delta\vev{\calo_\Delta}^2\, R^{2\Delta}$ is negligible in the limit $R\to0$. The remaining contributions in the first line then agree with the desired form \reef{tedsproposal} required for Jacobson's construction.

\subsection{Alternate quantization}
\label{alternate}

The holographic result presented in eq.~(\ref{deltaS}) is valid for any generic value of the scaling dimension $\Delta$. That is, eq. (\ref{deltaS}) is not only valid for $d/2<\Delta<d$ but also for $(d-2)/2<\Delta<d/2$, where $\Delta=(d-2)/2$ is the limit set by unitarity constraints.\footnote{As we show in the next section, $\Delta=d/2$ is a special case requiring a separate treatment.} However, as noted above, our translation of eq.~\reef{deltaS} to field theory quantities in eq.~\reef{jump} has only been justified for the first range of $\Delta$. In the regime $(d-2)/2<\Delta<d/2$, we need a different holographic renormalization procedure, that goes under the name of alternative quantization \cite{alter}. This is because for $\Delta<d/2$, we cannot longer have the usual relation between $\Delta$ and the mass of the scalar field as in eq. (\ref{mass}). Instead, we need
\begin{eqnarray}
\Delta = \frac{d}{2} - \sqrt{\frac{d^2}{4} + m^2 L^2}\,.
\labell{mass2}
\end{eqnarray}
Note that with this choice the roles of the normalizable and non-normalizable modes will be interchanged. That is, the scalar field still has the asymptotic expansion in \reef{such2}, but now the mode corresponding to $\phi_0$, which is still dual to the coupling $\lambda$, actually decays more rapidly in the limit $z\to0$.

Even though the need of an alternate quantization approach was pointed out in an early discussion \cite{alter} of the AdS/CFT correspondence, the computation of renormalized one-point functions using this approach appears not to have been carried in detail. Interesting ideas on how to construct a well-defined action in this regime and beyond the unitary bound were analyzed in \cite{don1,don2}.
In this section, we comment on how this procedure is performed and will present the main results coming from the alternate holographic renormalization. We leave the details and discussion of this method to Appendix \ref{alt_holo_ren} --- see also \cite{appear}.

By requiring that the expectation values are finite and that the Ward identities hold, we find the following unique regulated action for our case,
\begin{eqnarray}
I_{reg} = I_{bulk} + I_{GHBY} + I_{ct} + I_{Legendre}\,, \labell{fullaction}
\end{eqnarray}
where the first two terms are the standard contributions given in eqs.~\reef{ibulk} and \reef{boundary} and the last two boundary terms are given by
\begin{eqnarray}
I_{ct} & = & \left.-\frac{1}{2 \lp^{d-1}} \int d^dx \sqrt{-\gamma} \left( 2 (d-1) + \frac{\Delta}{2} \Phi^2 \right) \right|_{z=z_\epsilon} \,, \label{ct_alter} \\
I_{Legendre} & = & -\frac{1}{2 \lp^{d-1}} \int \left. d^dx \sqrt{-\gamma} \, \left(\Phi \ \hat{n}\! \cdot\! \nabla 
 \Phi -\Delta\,\Phi^2\right)\right|_{z=z_\epsilon} \,. \labell{legendreaction}
\end{eqnarray}
Note that in the counterterm action \reef{ct_alter}, the coefficient of $\Phi^2$ is modified here compared to the expression in eq.~\reef{boundary}. With this choice, the sum $I_{bulk} + I_{GHBY} + I_{ct}$ is finite, however, it does not satisfy the standard Ward identities, \ie the conformal and diffeomorphism Ward identities. The latter are restored by the addition of $I_{Legendre}$. In this last expression, $\hat{n}$ is the unit normal outgoing vector to the surface with $z=z_\epsilon$. Hence this term does not have the usual form of the counterterms considered in the holographic renormalization procedure. Instead, it should be considered as the term required for the Legendre transformation between the effective action and the generating functional \cite{alter} --- see further discussion in Appendix \ref{alt_holo_ren}.

Using the above action \reef{fullaction}, we find that the renormalized expectation values for $(d-2)/2<\Delta<d/2$ take exactly the same form given in eqs.~(\ref{stressed}--\ref{vevO}) for $d/2<\Delta<d$. Hence, the translation of eq.~\reef{deltaS} to field theory quantities is unchanged and $\delta S$ is given by precisely the same expression as before, namely, eq.~(\ref{jump}).
However, in this regime with $\Delta<d/2$, the contribution proportional to $\delta \vev{{\cal O}_\Delta}^2\,R^{2\Delta}$ dominates over the $R^d$ terms  involving the stress tensor. Therefore, these results do not agree with the desired form \reef{tedsproposal} for $\delta S_\mt{IR}$ and they seem to present a challenge for Jacobson's derivation of Einstein's equations \cite{ted,ted2}. We will return to discuss this point in section \ref{discuss}.

\subsection{$\Delta=d/2$}
\label{deltaover2}
It is quite clear that the expression for $\delta S$ in eq.~(\ref{formula}) does not apply for $\Delta=d/2$. In particular,  the coefficient of $\delta\vev{T^a{}_a}$ diverges with $\Delta\to0$. From the holographic perspective, the problem arises because the asymptotic expansion of the bulk scalar in eq.~\reef{such2} contains a single power of $z$ for both $\phi_1$ and $\phi_0$ when $\Delta=d/2$. Of course, the correct expansion takes the form 
\begin{eqnarray}
\Phi (z) =\ve \, z^{d/2} \left( \phi_0 \log(\mu z) + \phi_1 \right) + O(\ve^2)  \labell{such3}
\end{eqnarray}
to leading order in $\ve$. Note that we need to introduce an additional scale $\mu$ (with units of mass) to make the argument of the logarithm dimensionless. While this scale is rather arbitrary in the asymptotic expansion here, we can expect that it would be determined by infrared physics if we had a complete model of the holographic RG flow. Note that making a different choice of $\mu$ will change the value of $\phi_1$, which is still dual to $\vev{\cO_\Delta}$ --- see holographic renormalization below. The ansatz for the metric needs to change accordingly,
\begin{eqnarray}
g_{ab}(z) & = & \eta_{ab} + \ve^2 \, z^d \left( m_{ab} +b_{ab} \, \phi_1^2 + c_{ab} \, \phi_1 \, \phi_0 \, \log (\mu z) + d_{ab} \, \phi_0^2 \, \log^2(\mu z) \right) + O(\ve^3) \,, \nnn \\
\labell{wok}
\end{eqnarray} 
%
where $m_{ab}$, $b_{ab}$, $c_{ab}$ and $d_{ab}$ are matrices with constants to be determined by the Einstein's equations. This ansatz is motivated by the form of the scalar field expansion \reef{such3}, \ie every factor $\phi_0$ is accompanied by a $\log(\mu z)$. 

Now we solve eqs.~(\ref{einstein1}--\ref{KGeq2}) order by order in powers of $z$ (and $\log(\mu z)$) with $m^2 L^2 = -d^2/4$ (and $\kappa=0$). Here we use a redundancy in the parameterization of the metric function in eq.~\reef{wok} to simplify the following results. We obtain 
\begin{eqnarray}
g_{ab}(z) & = & \eta_{ab} + \ve ^2 z^d \left( m_{ab}  -  \frac{ \left(\phi_0 \log \left(\mu z\right)+\phi_1\right)^2}{4 (d-1)} \eta_{ab} \right) \,,
\end{eqnarray}
%
with the trace of the matrix $m_{ab}$ given by $m^a{}_a = \frac{\phi_0^2}{2 d (d-1)}$. 

Next step is to compute the area functional with the given metric to get $S_1$ and the same with vacuum expectation values for $S_0$. As in section \ref{eec}, at  order
$\ve^2$, we continue to use the hemisphere found in the AdS vacuum as the extremal surface and we just need to evaluate the area functional on this surface with the new metric. Then expanding the integrand in eq.~\reef{area8} to order $\ve^2$ and considering the desired difference of entropies, we find, after doing the angular integrals,
\begin{eqnarray}
\delta S = S_1 - S_0 = \frac{2\pi \, \Omega_{d-2}}{\lp^{d-1}}\ \ve^2 \int^R_0  \, \Delta s (z) \, dz \, ,
\end{eqnarray}
with
\begin{eqnarray}
\Delta s(z) = -\frac{z \left(R^2-z^2\right)^{\frac{d-3}{2}} \left((d-2) R^2+z^2\right) \left(2 \phi_0 \, \delta\phi_1 \log \left(\mu z\right)+\delta\phi_1^2-4 \delta m_{00} \right)}{8 (d-1) R} \,.
\end{eqnarray}
The remaining radial integral is finite and so there is no need to introduce a UV cut-off. Moreover, the integration can be performed analytically yielding 
\begin{eqnarray}
\delta S  =  \frac{2\pi \, \Omega_{d-2}}{\lp^{d-1}} \frac{R^d \ve^2}{d^2-1} & & \left(\frac{d \, \delta m_{00} }{2}- \frac{d}{4}\phi_0 \delta\phi_1 \left(\frac1{d (d+1)}-\frac12 H_{\frac{d-1}{2}}+ \log \left(\mu R\right)\right)-\frac{d \, \delta(\phi_1^2)}{8 } \right) \,, \nnn \\
\labell{deltaSwlog}
\end{eqnarray}
where $H_{\frac{d-1}{2}}$ is the harmonic number\footnote{Note that for odd values of $d$ (\ie integer $n$), this integral reduces to $H_n =\sum_{i=1}^n \frac{1}{i}$.} defined by $H_n =\int_0^1 dx \frac{1-x^n}{1-x}$. Note that there is an additional term proportional to $R^d \log(\mu R)$, which comes entirely from the logarithmic term already present in $\Delta s$. 



Next step is to write the expression in terms of the quantities in the boundary field theory. For that, we would like to obtain the renormalized expectation value for both the stress tensor and the operator of dimension $\Delta=d/2$. For this special case, we will need to introduce extra counterterms, \eg as shown in \cite{luis}. To eliminate the extra divergences arising from the logarithmic expansion of the scalar field, the counterterm action becomes
\begin{eqnarray}
I_{ct} & = & - \frac{1}{2\lp^{d-1}} \int d^d x \left. \sqrt{-\gamma} \left( 2(d-1) + \frac{d}{4} \Phi^2+\frac{1}{2 \log (\mu z)} \Phi^2 \right) \right\rvert_{z=z_\epsilon} \, .
\end{eqnarray}
This boundary term removes all of the divergences in the expectation values to order $\ve^2$. Note that the logarithmic term will also introduce some renormalization ambiguities in the definition of the expectation values, \eg see \cite{luis}. However, as we are interested in differences of expectation values, this will not be an issue here since the ambiguities cancel out in the subtraction.
Following a process analogous to that described in section \ref{holren} but now with the above counterterm action, we obtain
\begin{eqnarray}
\delta \vev{T_{ab}} & = & \frac{\ve^2}{2 \lp^{d-1}} \left( d \, \delta m_{ab} - \frac{1}{2}\,\eta_{ab}\, \phi_0 \, \delta \phi_1 \right) \, , \\
\delta \vev{T^a{}_a} & = & - \frac{\ve^2}{2 \lp^{d-1}} \frac{d}{2} \phi_0 \delta \phi_1  \,,\\
\vev{\calo_\Delta} & = & - \frac{\ve}{2 \lp^{d-1}} \, \phi_1 \,.
\end{eqnarray}

Given these results, as well as eq.~\reef{relations}, we can now write the variation in the entanglement entropy as
\begin{eqnarray}
\delta S &=& \frac{\Omega_{d-2} R^d}{d^2-1} \left( \delta \vev{T_{00}} + \delta \vev{T^a{}_a} \left( \frac{d+2}{d(d+1)} - \frac{1}{2} H_{\frac{d-1}{2}}+\log (\mu R) \right)\right)\nonumber\\
&&\qquad\qquad\qquad\qquad  -\frac{2^{d-1}d^2}{d-1}\, \frac{R^d\,\delta \vev{\calo_\Delta}^2}{C_T}\,. \labell{gush}
\end{eqnarray}
Hence in the small $R$ expansion, the leading term is proportional to $R^d \log(\mu R)$ and the desired $R^d$ contribution is actually a subdominant contribution. Hence, we again find that our holographic results here are in conflict with the form assumed in eq.~\reef{tedsproposal}. We will return to discuss this point in section \ref{discuss}.

\subsubsection{Massive Dirac Fermions in $d=2$} \label{mdf}

It is interesting that with precisely $\Delta=d/2$, $\delta S$ acquires a logarithmic term in our holographic calculation above.  It is important to understand the appearance of this term is special to such a holographic framework or if similar terms arise with general CFT's. In the case of the free massive Dirac fermion in $d=2$, the modular Hamiltonian is known exactly and so can be computed perturbatively for small mass \cite{massive2,massive3}. In this case, the mass operator has dimension $\Delta=d-1=1=d/2$ and so it is possible to check whether the logarithmic behaviour is also present in that context.

We will expand the modular Hamiltonian of a Dirac field with mass $m$ on an interval of size $2R$ in $d=2$ for small $m R$. 
The modular Hamiltonian for one interval is
\begin{equation}
{\cal H}=\int_{-R}^R dx\, dy\, \Psi^\dagger(x) H(x,y) \Psi(y)\,.
\end{equation}
The kernel $H(x,y)$ is given by
\begin{equation}
H=-\int_{1/2}^\infty d\beta\, (R(\beta)+R(-\beta))\,,\labell{dual}
\end{equation}
in terms of the resolvent
\begin{equation}
R(\beta)=(C-1/2+\beta)^{-1}\,,
\end{equation}
where $C(x,y)=\langle 0|\psi(x)\psi^\dagger(y)|0\rangle$ is the correlator kernel in the interval. Expanding $C$ to first order in the mass we have
\begin{equation}
R(\beta)=R^0(\beta)-R^0(\beta)\,\delta C \,R^0(\beta)+\cdots\labell{23}
\end{equation}
with \cite{massive2}
\bea
R^0(\beta)(x,y)&=&\int_{-\infty}^\infty ds\, \psi_s(x) M(\beta,s) \psi^*_s(y)\,,\\
M(\beta,s)&=&\left(\beta {\bf 1}-\tanh(\pi s)\frac{\gamma^3}{2}\right)^{-1}\,,\\
\psi_s(x)&=& \frac{R^{1/2}}{\pi^{1/2}\sqrt{R^2-x^2}}e^{-i s z(x)}\,,\\
z(x)&=&\log\left(\frac{R+x}{R-x}\right)\,,\\
\delta C(x,y)&=&-\frac{m}{2\pi}\left(\gamma_E-\log(2)+\log(m|x-y|)\right)\gamma^0\,.
\eea
Here $\gamma^0,\gamma^1$ are the Dirac matrices and $\gamma^3=\gamma^0\gamma^1$.

The zeroth order calculation gives the expected conformal result \cite{massive2} --- see also eq.~\reef{modu} below:
\begin{equation}
{\cal H}_0=2\pi\int_{-R}^R dx\,  \frac{R^2-x^2}{2R} (i/2) \Psi^\dagger \gamma^3 \stackrel{\leftrightarrow}{\partial}_x \Psi\,.\labell{10}
\end{equation}

To compute the first order in the mass, since we are interested in the small size limit of the expectation values of ${\cal H}$, we can replace
\begin{equation}
\langle {\cal H}_1\rangle \sim  \left(\int_{-R}^R dx\,dy\, H_1(x,y)\right)\,\,\langle\bar{\Psi}\Psi\rangle=K \langle\bar{\Psi}\Psi\rangle \,, \labell{kkx}
\end{equation}
where we have used that the first order contribution is proportional to the Dirac $\gamma^0$ matrix. That is, we only need the kernel $H_1$ integrated in the interval. A more detailed calculation of local and non-local terms in $H_1$ will 
be presented in \cite{dos}.

We insert the second term of (\ref{23}) into (\ref{dual}), do the integral in $\beta$ in (\ref{dual}), and use  
\begin{equation}
\int_0^L dx\, \psi_s(x)=\left(\frac{\pi}{2}\right)^{1/2} L^{1/2} \textrm{sech}(\pi s)\labell{use}
\end{equation}
to do the integrals in $x,y$ in (\ref{kkx}). We find
\begin{eqnarray}
K&=&-2\pi R m \int_{-\infty}^\infty ds\,ds^\prime\, \frac{(s+s^\prime)}{\sinh(s+s^\prime)} \\
&& \hspace{1.5cm} \int_{-R}^R dx\, \int_{-R}^R dy\,  \psi^*_s(x)\left(\gamma_E-\log(2)+\log(m L) +\log(|x-y|/L)\right) \psi_{s^\prime}(y) \,. \nonumber
\end{eqnarray}
The constant term in the brackets can be integrated using (\ref{use}). The integral of the logarithmic term can be obtained passing to the variables $u=s+s^\prime$, $v=s-s^\prime$ and doing the integrals over $u,v$ first, and then the integrals over $x,y$. We finally get
\begin{equation}
\langle {\cal H}_1\rangle= - \frac{4\pi}{3} m R^2 \left(\alpha+\gamma_E+\log(m R) \right)\langle\bar{\Psi}\Psi\rangle\,.
\end{equation}
with 
\begin{equation}
\alpha=\frac{3 \pi^2}{2}\int_0^1 dx\, \frac{\log|1-2 x|}{1+\cosh(\pi \log(x/(1-x)) )}\simeq-4.53085\,.
\end{equation}

Together with the leading term (\ref{10}), this gives
\begin{equation}
\langle {\cal H}\rangle=\frac{4\pi R^2}{3} \left(\langle T_{00}\rangle   -  \left(1+\alpha+\gamma_E+\log(m R) \right)\vev{T^a{}_a}\right)\,, \labell{finalfermion}
\end{equation}
where we have used $T_{00}=(i/2) \Psi^\dagger \gamma^3\!\! \stackrel{\leftrightarrow}{\partial}_x\!\!\! \Psi+m \bar{\Psi} \Psi$ and $T^a{}_a=m \bar{\Psi}\Psi$. Of course, $\gamma_E$ denotes the Euler-Macheroni constant,
\ie $\gamma_E\simeq0.5772157$. 

Now if consider applying the first law of entanglement \reef{firstlaw}, we arrive at
 \begin{equation}
\delta S=\delta\langle {\cal H}\rangle=\frac{4\pi R^2}{3} \left(\delta\langle T_{00}\rangle   -  \left(1+\alpha+\gamma_E+\log(m R) \right)\delta\vev{T^a{}_a}\right)\,, \labell{finalfermion2}
\end{equation}
which exactly coincides with the holographic result (\ref{formula2}), except for the constant coefficient multiplying $\vev{T^a{}_a}$.\footnote{Above, we have the coefficient
$1+\alpha+\gamma_E\simeq-2.95$ for the free fermion, while substituting $d=2$ into the holographic result \reef{formula2}, the corresponding constant is $\frac23-\frac12 H_{1/2}\simeq -0.0323$.} In order to properly compare this coefficient, a physical choice for the mass scale $\mu$ in (\ref{formula2}) must first be fixed in a more complete holographic model. In any event, we may conclude that the logarithmic contribution is not an artifact of the holographic calculations.

\section{Discussion} \label{discuss}

Jacobson's derivation \cite{ted,ted2} of Einstein's equations makes a precise connection between entanglement and gravity. However, his argument relies on two key assumptions: The first is that the entanglement entropy for the vacuum reduced to a small ball is maximal for variations holding the volume fixed. We have nothing to add on this point in this paper and will simply accept this postulate in the following discussion. The second assumption is that the variation of the entanglement entropy coming from variation of the matter fields takes a certain form given in eq.~\reef{tedsproposal}. This is the assumption that we examined in detail here for a class of holographic models. In particular, our holographic calculations  evaluated the variation of the entanglement entropy for small spheres where the boundary theory is deformed by a relevant operator ${\cal O}_\Delta$. 

Let us begin with two technical comments on our results in eqs.~\reef{formula} and \reef{formula2}: First, the variation of the entropy $\delta S_\mt{IR}$ is only a scalar quantity but ref.~\cite{ted} is deriving the full tensor comprising Einstein's equations \reef{Einstein}. Hence it is important that the variation has the form $\delta S_\mt{IR}= Y_{ab}\,\hat{t}^a \hat{t}^b$ where $Y_{ab}$ is some symmetric tensor and $\hat{t}^a$ is the unit time-like vector orthogonal to the Cauchy slice containing the spherical region for which we are evaluating the entanglement entropy. Since our analysis was done for a general state (\ie state which is anisotropic and stationary), it is straightforward to verify that our holographic results for $\delta S_\mt{IR}$ take this form. In particular, we can boost any given background to a new reference frame while leaving the entangling sphere fixed, and the form of our results is unchanged. That is, the first contribution in either eq.~\reef{formula} or \reef{formula2} is proportional to $\delta\vev{T_{00}}$ in the new frame and hence corresponds to a term proportional to $\delta\vev{T_{ab}}$ in $Y_{ab}$. Similarly, the contributions proportional to  $\delta\vev{T^a{}_a}$ and $\delta\vev{\cO_\Delta}^2$ are left unchanged and so we can interpret these two terms as appearing in $Y_{ab}$ with a factor of $-g_{ab}$.

Our second technical comment has to do with the factor of $1/C_T$ appearing in the contributions proportional to  $\delta\vev{\cO_\Delta}^2$. Naively, one may think that this factor indicates that this term is suppressed relative to the others because our holographic framework requires that we are working with a large central charge, \eg this is the usual large $N$ limit for holographic gauge theories. However, our normalization is such that the expectation values of any (single-trace) operators are themselves proportional to $C_T$, \eg as revealed by the factor of $1/\lp^{d-1}$ appearing in eqs.~(\ref{stressed}--\ref{vevO}). Hence this factor of $1/C_T$ ensures that all three contributions in eqs.~\reef{formula} and \reef{formula2} are contributing at the same order in this regard, \ie they are all proportional to $C_T$. A short calculation in Appendix \ref{slider} of $\delta S_\mt{IR}$ for a thermal state in a general $d=2$ CFT shows that the appearance of such factors is natural in evaluating entanglement entropy for CFTs, even beyond holography.

Our results in eqs.~\reef{formula} and \reef{formula2} found the appearance of a contribution to $\delta S_\mt{IR}$ proportional to $R^{2\Delta}\,\delta\vev{{\cal O}_\Delta}^{2}$, where the power of $R$ is completely fixed on dimensional grounds. Such a contribution seems problematic in the regime $(d-2)/2 < \Delta < d/2$ since this term decays more slowly  than the desired $R^d$ terms in the limit $R\to0$.\footnote{In Appendix \ref{higher}, we also identified a contribution proportional to $R^{3\Delta}\,\delta\vev{{\cal O}_\Delta}^{3}$, which also becomes problematic for $\frac{d-2}2<\Delta<\frac{d}3$ with $d<6$. However, this contribution will always be subdominant compared to the term discussed above in the main text.} Hence it is natural to think that $\delta S_\mt{IR}$ would be dominated by this contribution and so Jacobson's argument, which relies on the form \reef{tedsproposal}, would be invalid. One obvious resolution of this problem would be if there were no such operators in the UV fixed point theory that describes the matter fields in our universe. Again, the unitarity bound for general CFT's allows (scalar) operators with conformal dimensions in this problematic regime.\footnote{In fact, explicit examples are known in certain supersymmetric gauge theories, \eg see \cite{alter}.} Hence this requirement would be a restriction on the spectrum of operators appearing in the matter sector of a theory described Einstein gravity.

Let us also observe that the expectation values appearing in eq.~\reef{formula} are all set by infrared physics scales which are independent of the size of the sphere $R$. Now in general, we would have
\beq
\delta\,\vev{\cO_\Delta}^2\simeq C_T^2\,\mu_\cO^{2\Delta}\,,\quad
 \delta\vev{T_{00}}\simeq C_T\,\mu_0^d \quad
{\rm and}\quad \delta\vev{T^a{}_a}\simeq C_T\,\mu_T^d\,.
\label{skales}
\eeq
Of course, we are considering the regime where all of these energy scales are much smaller than that set by the radius of the sphere, \ie $\mu_{\cO,0,T}\ll1/R$.
However, in arriving at the conclusion that the $\delta\,\vev{{\cal O}_\Delta}^{2}$ contribution creates a problem, we are implicitly assuming that these scales are all roughly the same, \ie $\mu_\cO\simeq\mu_0\simeq\mu_T \simeq \mu$, so that the different contributions in eq.~\reef{formula} can be compared with powers of the same dimensionless product  $\mu R$. However, in general, there is no need for these scales to be the same. In particular, one can imagine that there will be broad families of states where $\mu_\cO\ll \mu_{0,T} $. Then even if there are operators with $\Delta<d/2$, one may still have
$(\mu_\cO R)^{2\Delta}\ll (\mu_{0,T}R)^{d}\ll1$ for small but finite $R$ in a broad class of states. In this case, gravity would be properly described by Einstein's equations in this family of states but it raises the intriguing possibility that this description would breakdown in other `low entropy' states.\footnote{As we will see below, the $\delta\vev{{\cal O}_\Delta}^{2}$ contribution tends to make $\delta S_\mt{IR}$ smaller than required for Jacobson's derivation.} We return to discussing this possibility later in this section.

We also considered the special case of $\Delta=d/2$ for which $\delta S_\mt{IR}$ is given by eq.~\reef{formula2}. In this case, the term proportional to $\delta\vev{{\cal O}_\Delta}^{2}$ carries a factor of $R^d$ and so this contribution can simply be absorbed into the $\vev{X}$ term in eq.~\reef{tedsproposal}. However, there is a additional contribution proportional to $\delta\vev{T^a{}_a}$ with a factor of $R^d\,\log(\mu R)$. In fact, in section \ref{mdf}, we confirmed the appearance of this extra logarithmic dependence beyond the framework of holography. There, we found that that the same term appears for a free Dirac fermion in two dimensions, for which the entanglement Hamiltonian is explicitly known \cite{massive2,massive3}. When the theory is perturbed by a small mass, \ie $mR\ll1$, it is quite remarkable that the same logarithmic term appears in eq.~\reef{finalfermion} with the precisely same coefficient as in the holographic result \reef{formula2}.

The extra logarithm gives an enhanced, although only mildly enhanced, dependence on the radius $R$ so that the appearance of this term in $\delta S_\mt{IR}$ is again problematic for Jacobson's construction. Since the appearance of this term requires a precise value for the conformal dimension, \ie  $\Delta=d/2$, it may seem more reasonable to require no such operators appear in the UV fixed point theory of the matter fields. Of course, in four dimensions, a mass term for a free scalar field  would be a canonical example of such a term. However, one should expect that unless the scalar is completely free that even weak interactions will induce a small anomalous dimension and hence eliminate the appearance of this problematic contribution to $\delta S_\mt{IR}$.

We should emphasize that, apart from section \ref{mdf}, the calculations here are limited to holographic theories with an Einstein gravity dual. Hence there are the usual caveats that the corresponding CFTs should have a large central charge, be strongly coupled and have a sparse spectrum. A priori, it is not clear how universal the results obtained here would be for more general theories. However, it seems that in fact our calculations may extend to generic CFTs following the approach of \cite{deff}. The latter reference  argued that when a generic CFT is deformed as in eq.~\reef{perturb}, $\delta S_\mt{IR}$ is completely determined by universal two- and three-point correlators in the CFT and further that the result could be evaluated by recasting it into the form of a holographic gravity calculation similar to those presented here. Note, however, that ref.~\cite{deff} evaluated the change in the entanglement entropy between the CFT vacuum and the vacuum of the deformed theory. Hence it remains to consider excited states in the deformed theory --- see \cite{rough}. However, the primary challenge is to extend these calculations to conformal dimensions in the regime found to be of most interest here, \ie $\frac{d-2}2<\Delta\le\frac{d}2$.

\subsection*{Modular Hamiltonians}

As described in the introduction, an important contribution to the variation of the entropy comes from the first law of entanglement \reef{firstlaw}. In the present discussion, examining the expectation value $\langle {\cal H}\rangle$ for small spheres is related to examining what would be commonly referred to as the `operator product expansion' (OPE) of the modular Hamiltonian. That is, $\cal H$ would be given by some complex and generally nonlocal expression involving collections of operators restricted in a finite region, \ie the sphere of radius $R$. However, if it is only examined with long wavelength probes, we can effectively approximate $\cal H$ by a sum of local operators.\footnote{The interested reader can find more detailed considerations of the OPE for Wilson lines and surface operators in gauge theories in \cite{shifman} and of twist operators in higher dimensional CFTs in \cite{twist}.} 

In fact, reducing the flat space vacuum of a CFT to a sphere of radius $R$ yields a remarkably simple expression for the modular Hamiltonian \cite{CHM}
\beq
{\cal H}=  2\pi \int_{r\le R} \!\!\!d^{d-1}x\  \frac{R^2-r^2}{2R}\ T_{00}\ +\ c'\,,
\labell{modu}
\eeq
where the constant $c'$ is fixed by demanding that the corresponding density matrix is normalized with unit trace. Hence in this case, the OPE only involves the energy density and its derivatives, but of course, the derivative terms are accompanied higher powers of $R$, \ie
\beq
{\cal H}\simeq \frac{2\pi\,\Omega_{d-2}}{d^2-1}\,R^d\left[ \, T_{00} + \frac{1}{2(d+3)}\, R^{2}\, \nabla^2 T_{00} +\cdots\right]\,.
\labell{OPE}
\eeq
Hence in these additional terms are higher order contributions in the limit of small $R$, which are then negligible for the purposes of Jacobson's argument. 

Let us add that Jacobson's general argument compares a given state to the vacuum in a maximally symmetric background, \ie Minkowski space, de Sitter space or anti-de Sitter space. This approach allows him to accommodate the possibility of a cosmological constant, as well as the scalar contribution $\vev{X}$ from nonconformal matter fields in eq.~\reef{tedsproposal}. One can easily extend the construction of \cite{CHM} to evaluate the modular Hamiltonian for a CFT in the dS or  AdS backgrounds. For example, let us consider the static patch of dS space, \ie
\beq
ds^2=-f^2(r)\,dt^2+\frac{dr^2}{f^2(r)}+
r^2d\Omega^2_{d-2}\quad{\rm with}\ \ f^2(r)=1-\frac{r^2}{L^2}\,.
\labell{sitter}
\eeq
For a spherical region of radius $R$ placed at the origin in the above coordinates, the modular Hamiltonian becomes
\beq
{\cal H}_\mt{dS}=  2\pi \int_{r\le R} \!\!\!\!\!d\Omega\,dr\,r^{d-2}\  \frac{ L^2}{R}\ \frac{f(r)-f(R)}{f^3(r)} \ T_{00}\ +\ c''\,,
\labell{modu2}
\eeq
Of course, the modular Hamiltonian \reef{modu2} is still given by a local integral of the energy density alone. We are interested in the regime where the radius of the sphere $R$ is much smaller than the dS curvature scale $L$.\footnote{We have chosen coordinates such that with $L\to\infty$, eq.~\reef{modu2} reduces to the standard flat space expression \reef{modu}. Further, the area of the spherical entangling surface is $\Omega_{d-2}R^{d-2}$, independent of $L$.} Then the leading curvature correction in the OPE expansion is
\beq
{\cal H}_\mt{dS} \simeq \frac{2\pi\,\Omega_{d-2}}{d^2-1}\,R^d\left[ \, T_{00} + \frac{2d-1}{d+3} \, \frac{R^{2}}{L^2}\, T_{00} +\cdots\right]\,.
\labell{OPE2}
\eeq
That is, the modifications due to the curvature scale in the first law \reef{firstlaw} are suppressed by powers of $R/L\ll1$ and make a negligible contribution in Jacobson's construction. We expect that curvature contributions will again be suppressed in a similar way for the case of a deformed CFT.

Given eq.~\reef{modu}, one may conclude that 
for a CFT, the contributions proportional to $\vev{{\cal O}_\Delta}^{2}$ appearing in eqs.~\reef{formula} and \reef{formula2} take us beyond the first law of entanglement \reef{firstlaw}. The coupling $\lambda$ does not explicitly appear in these terms and so they would also appear for excited states of the CFT, \ie even when $\lambda=0$. However, as noted above, all of the contributions from the modular Hamiltonian in a CFT will only involve the energy density and its derivatives. Hence the $\vev{{\cal O}_\Delta}^{2}$ contribution cannot be contained in the expression for $\delta\vev{\cal H}$ on the right-hand side of eq.~\reef{firstlaw}. In this sense, these must be `higher order' contributions to $\delta S_\mt{IR}$ that go beyond the first law. At this point, we should stress that Jacobson's arguments only considered first law contributions to $\delta S_\mt{IR}$ --- we return to this point in the discussion below. 
Further, we observe that in the case of a CFT, we have $\vev{{\cal O}_\Delta}_{\rm vacuum}=0$ and so
\beq
\delta\,\vev{{\cal O}_\Delta}^{2}
=\vev{{\cal O}_\Delta}^2\ge0 \,.\labell{corn}
\eeq
 Hence the sign of the corresponding coefficient in eqs.~\reef{formula} and \reef{formula2} is such that $\delta S_\mt{IR} \le  \delta\vev{\cal H}$ as required by the positivity of relative entropy, \eg \cite{relative}. 
 
A similar discussion applies for a deformed CFT, where we may expect that $\langle {\cal O}_{\Delta}\rangle_{\rm vacuum}\ne0$. In this case, we may write
 \beq
 \delta\,\vev{{\cal O}_\Delta}^{2}
=\big(\delta\vev{{\cal O}_\Delta}\big)^2+2\, \langle {\cal O}_{\Delta}\rangle_{\rm vacuum}\ \delta\langle {\cal O}_{\Delta}\rangle  \,,\labell{corn11}
\eeq
where $\delta\vev{\cO_{\Delta}}=\vev{\cO_{\Delta}}-\langle {\cal O}_{\Delta}\rangle_{\rm vacuum}$. 
Now the second term on the right-hand side is linear in the deviation of $\vev{\cO_{\Delta}}$ away from the deformed vacuum. Hence this contribution must come from the variation of the expectation value of the modular Hamiltonian (of the deformed CFT) in the first law.\footnote{Further, the coefficient of this variation is some function of the coupling, \ie $\langle {\cal O}_{\Delta}\rangle_{\rm vacuum}=f(\lambda)$ with $f(\lambda=0)=0$. Hence, we can think of this as a building block available in construction of the the modular Hamiltonian of the deformed theory.} Note that this second term above can have either sign. In contrast, the first term in eq.~\reef{corn11} is positive, being quadratic in the deviation from the deformed vacuum. Hence this contribution goes beyond the first law again and gives the leading term in the relative entropy with $\delta \langle {\cal H}\rangle-\delta S\ge0$.   

Further, let us add that the contribution to $\delta S_\mt{IR}$ proportional to the trace of the stress tensor $\delta\vev{T^a{}_a}$ also comes from the modular Hamiltonian, since again it is linear in the deviation of the expectation value from that in the deformed vacuum. Our assertion is also supported by results in section \ref{mdf} for massive fermions in two dimensions, which only considers the first law contribution. Further support comes from the results in \cite{deff}, which indicate that simply replacing $T_{00}$ by $T_{00} -g_{00}\,T^a{}_a/(2\Delta-d)$ in eq.~\reef{modu} yields the modular Hamiltonian of the deformed CFT to leading order in a $\lambda$ expansion.\footnote{We expect that if we evaluate the modular Hamiltonian for any Cauchy slice other than $t=0$, it will involve a nonlocal expression even at first order in $\lambda$.} This result would apply for $d/2<\Delta<d$ in generic CFTs and would yield precisely the first line of our result for $\delta S_\mt{IR}$ in eq.~\reef{formula}. Note however that this simple expression for the modular Hamiltonian is singular for $\Delta=d/2$ and so will not apply for this value of the conformal dimension. Our $d=2$ fermion calculations in section \ref{mdf}, which apply to this case, emphasize that the local expressions such as those in eq.~\reef{formula} or \reef{formula2} emerge from taking the OPE limit of the modular  Hamiltonian. That is, for the massive free fermions, the modular Hamiltonian is a nonlocal expression in general, even to leading order in the mass deformation. 

In general then, the OPE limit will yield $\delta S_\mt{IR}$ as a sum of expectation values of local operators, plus expectation values of local operators squared and higher powers.  As in eq.~\reef{OPE}, the coefficients of higher dimension operators or higher powers of expectation values will include higher powers of the radius $R$ and so these tend to give subleading terms. This structure will be the completely general independent of holography or the particular details of the theory under study. Moreover, one may ask which of the contributions in this expansion are determined by the modular Hamiltonian, \ie the first law \reef{firstlaw}, and which are not? In general, the expectation value of modular Hamiltonian will yield contributions which are linear in the excitation of the expectation value of any operator above its vacuum expectation value (\eg $\delta\vev{T_{00}}$, $\delta\vev{T^a{}_a}$ or $\delta\vev{\cO_\Delta}$). Any contributions which are not linear in such expectation values extend $\delta S_\mt{IR}$ beyond the first law.

\subsection*{Four Roads to Quantum Gravity} 

At this point, we would like to assess the implications of our results for Jacobson's derivation of Einstein's equations \cite{ted,ted2}. It seems that there are at least four different possible interpretations:\footnote{We would like to thank Ted Jacobson for his suggestions and comments on the following.}

The first and most straightforward seems to be that the derivation only applies to linear variations around the vacuum. That is, Jacobson's original argument only considered variations of the entanglement entropy consistent with the first law \reef{firstlaw}, whereas we showed that the problematic term proportional to $R^{2\Delta}\delta\vev{\cO_\Delta}^2$ (with $\Delta<d/2$) in eq.~\reef{formula} extends $\delta S_\mt{IR}$ beyond this range. Hence, the correct interpretation of Jacobson's derivation may be to restrict attention to linear variations about the vacuum. As pointed out in \cite{ted}, this interpretation already seems to be necessary when considering coherent states, which can have a finite energy density while leaving the entanglement entropy unchanged \cite{one,two}.\footnote{These coherent states are not in conflict with the first law \reef{firstlaw} because the energy density is second order in the amplitude of the matter fields.} Unfortunately, in this case, the derivation would only reveal the {\it linearized} Einstein equations, similar to the holographic analysis in \cite{wow}. Moreover, it would be useful to investigate more thoroughly (\ie beyond holography) the possibility that low dimension terms may appear in the OPE expansion of the modular Hamiltonian to test the validity of this interpretation \cite{dos}. 

Above, we already suggested an alternate resolution of this issue. Namely, no problematic second order terms would arise if there were no $\Delta<d/2$ operators in the UV fixed point theory that describes the matter fields in our universe. That is, the spectrum of operators appearing in the ultraviolet would be restricted in order for the infrared theory to be described by Einstein gravity. In fact, the first approach would still require such a restriction to avoid the logarithmic contribution which appears in the first law when $\Delta=d/2$, as shown in eq.~\reef{formula2}. Ruling out only operators with precisely $\Delta=d/2$ in the UV fixed point theory would seem to be a more mild restriction. 

A third proposal (by Ted Jacobson) is that the contributions proportional to $R^{2\Delta}\delta\vev{\cO_\Delta}^2$ could be absorbed in a manner analogous to the treatment of $\delta\vev{X}$ in eq.~\reef{tedsproposal}.\footnote{The following approach could also be adapted to absorb the logarithmic contribution in eq.~\reef{formula2}.} In the latter case, the curvature of the maximally symmetric reference background is chosen locally to absorb this scalar expectation value, \ie
\beq
G^{MSS}_{ab}=-\tlam\,g_{ab}\qquad{\rm with}\quad \tlam=\Lambda+8\pi G\,\delta\vev{X}\,.
\labell{hammer}
\eeq
If we extend this choice to allow dependence on both the state and the size of the sphere, then setting
\beq
\tlam=\Lambda+8\pi G\,\delta\vev{X}+8\pi G\, \frac{k_\Delta}{C_T\,R^{d-2\Delta}}\,\delta \vev{\cO_\Delta}^2
\labell{hammer2}
\eeq
will absorb the second order term and produce the expected Einstein equations \reef{Einstein}. Here, the (positive) constant $k_\Delta$ is given by the ratio of the numerical coefficients of $\delta \vev{\cO_\Delta}^2$ and $\delta\vev{T_{00}}$ in eq.~\reef{formula}. This approach has the uncomfortable feature that $\tlam$ grows arbitrarily large as $R\to0$ (when $\Delta<d/2$). Note, however, that $C_T\, (\mu_\cO R)^{2\Delta}\, (\lp/R)^{d-2}\ll 1$ ensures that the radius of curvature set by $\tlam$ is still much larger than the radius of the sphere $R$.\footnote{Here $\lp^{d-2}\equiv 8\pi G$ (rather than the Planck scale in the holographic theory). Also recall that $\mu_\cO$ was defined in eq.~\reef{skales}. Of course, Jacobson's derivation requires both $\mu_\cO R\ll1$ and $\lp/R\ll 1$.} This interpretation would then to allow for the small but finite variations and the derivation of the full nonlinear Einstein equations. However, this approach still calls for a better understanding to the role of the maximally symmetric reference geometry in Jacobson's construction.

A last more speculative possibility  is that Jacobson's derivation indicates that particular class of states which `gravitate' according to Einstein's equations.
Recall that we suggested that the $R^{2\Delta}\delta\vev{\cO_\Delta}^2$ contribution would not be a problem, even with $\Delta<d/2$, as long as  $\mu_\cO\ll\mu_{0,T}$, where these are the scales characterizing the various expectation values in eq.~\reef{skales}. Hence, Jacobson's derivation still carries through and gravity would still be described by Einstein's equations for the family of states. However, this also raises the intriguing prospect that Einstein's equations will `fail' for low entropy states, where $\mu_\cO\gtrsim\mu_{0,T}$. The idea that Einstein's equations break down for highly excited states is not a surprising one. For example, when the stress energy reaches the Planck scale, \ie $\mu_{0,T}\sim1/\lp$, in the early universe, one no longer expects that the cosmological evolution is described by Einstein's equations. However, the interesting feature of the breakdown anticipated here is that none of the scales involved need to be Planckian. Instead, one can anticipate a failure of Einstein's equations even when the energy density is much much less than the Planck scale.

If we naively carry Jacobson's derivation ahead with eq.~\reef{formula} (and use eq.~\reef{hammer} to choose the reference geometry), the gravitational equation becomes
\beq
G_{ab} + \Lambda g_{ab} = 8 \pi G \, \left[ \vev{T_{ab}} + \frac{k_\Delta}{C_T\,R^{d-2\Delta}}\,g_{ab}\,\delta \vev{\cO_\Delta}^2+\cdots\right]\,,
\labell{Einstein2}
\eeq
where as in eq.~\reef{hammer2}, $k_\Delta$ corresponds to the ratio of the coefficients of $\delta \vev{\cO_\Delta}^2$ and $\delta\vev{T_{00}}$ in eq.~\reef{formula}. Hence the new $\delta \vev{\cO_\Delta}^2$ term might be seen as an additional contribution to the cosmological constant term in eq.~\reef{Einstein2}. However, to properly interpret this equation, it remains to understand the role of $R$. Of course, $R$ was the radius of the spherical entangling surface but this is simply an auxiliary scale in Jacobson's derivation that disappears from the final Einstein equation \reef{Einstein}. One suggestion would be that $R$ can be regarded roughly as a renormalization scale in eq.~\reef{Einstein2}. That is, we should think of eq.~\reef{Einstein2} as the appropriate gravitational equation when we are probing the spacetime on (length) scales of the order of $R$. Of course, this interpretation seems incomplete since we cannot choose $R$ arbitrarily rather the construction requires $\mu_{\cO,0,T} \ll 1/R$.

Further, let us comment that the sign of the coefficient $k_\Delta$ is fixed to be positive by the positivity of relative entropy, as commented below eq.~\reef{corn}. As further noted there, $\delta \vev{\cO_\Delta}^2$ is also guaranteed to be positive at a conformal fixed point and so the new contribution in eq.~\reef{Einstein2} would resemble to a negative energy density. Beyond a conformal fixed point, the sign of $\delta \vev{\cO_\Delta}^2$ might have either sign. 

Hence if we apply this interpretation of Jacobson's derivation, it seems that we are naturally led to conclude that Einstein's equations may `fail' for certain classes of low energy states. It will be interesting to better understand how gravity is modified in these states and the implications of these modifications. In particular, they may have important consequences for our understanding of the cosmological constant problem, early universe cosmology and perhaps warp drive engineering \cite{warp}.

\section*{Acknowledgements}
We would like to thank A. Buchel, J. McGreevy and C. Uhlemann for useful discussions. Further, we especially thank T. Jacobson and A. Speranza for interesting discussions and correspondence about our results, and for comments on the manuscript. Research at Perimeter Institute is supported by the
Government of Canada through the Department of Innovation, Science and Economic Development and by the Province of Ontario
through the Ministry of Research \& Innovation. RCM and DAG are also supported
by an NSERC Discovery grant. RCM is also supported by research funding
from the Canadian Institute for Advanced Research. HC and RCM also acknowledge support from the Simons Foundation through ``It from Qubit" Collaboration.
DAG also thanks the Kavli Institute for Theoretical Physics, where he was a Visiting Graduate Fellow, for hospitality and support during the first stages of this project. Research at KITP is supported, in part, by the National Science Foundation under Grant No. NSF PHY11-25915.

\appendix

\section{Alternate Holographic Renormalization}
\label{alt_holo_ren}

The aim of this Appendix is to show how to perform holographic renormalization for operators in the range $(d-2)/2 < \Delta <d/2$. In particular, we are interested in obtaining renormalized one-point functions of the operator and the stress energy tensor which: (a) are  UV finite and (b) satisfy both the conformal and the diffeomorphism Ward identities. In the usual regime with $\Delta>d/2$, the well-studied procedure of holographic renormalization gives the desired expectation values that satisfy both Ward identities. However, the extension to cover the $\Delta<d/2$ case is not straightforward and here we present a consistent way of doing so. For that, we will merge the ideas that first appeared in \cite{alter} with the modern approach of holographic renormalization. We also consider a slightly different approach considered in \cite{don2} to derive the same action.

In \cite{alter}, it was proposed that the generating functional of the theory in which $\Delta$ is given by eq. (\ref{mass2}) is in fact the Legendre transformation of the one with $\Delta$ given by eq. \reef{mass}. This would suggest that in the holographic renormalization procedure we should add a term that would play the role of this Legendre transformation. In fact, this will be a boundary term of the form $\Phi \, (\hat{n}\! \cdot\! \nabla) \, \Phi$, where $\hat{n}$ is the unit normal vector to the surface with fixed $z=z_\epsilon$. So the most general action that includes all possible counterterms that are relevant to our case\footnote{In general, we can have counterterms proportional to curvatures of the boundary metric, for instance. We are neglecting those terms are we are fixing a flat boundary metric. Other terms including higher powers of $\Phi$ are also negligible as they will be higher order in the expansion parameter $\ve$.}, and also includes this Legendre transformation term is
\begin{eqnarray}
I_{reg} = I_{bulk} + I_{GHBY} + I_{ct} + I_{Legendre}\,,
\end{eqnarray}
where the first two terms are the standard contributions given in eqs.~\reef{ibulk} and \reef{boundary} and the last two boundary terms are given by
\begin{eqnarray}
I_{ct} & = & \left.-\frac{1}{2 \lp^{d-1}} \int d^dx \sqrt{-\gamma} \left( 2 (d-1) + \frac{\Delta}{2} \Phi^2 \right) \right|_{z=z_\epsilon} \,, \label{ct_alter2} \\
I_{Legendre} & = & -\frac{1}{2 \lp^{d-1}} \int \left. d^dx \sqrt{-\gamma} \, \left( B \ \Phi \ \hat{n}\! \cdot\! \nabla \Phi + A\, \Phi^2\right)\right|_{z=z_\epsilon} \,. \labell{wacko}
\end{eqnarray}
Here, $\gamma$ is the determinant of the induced metric in the surface $z=z_\epsilon$, and $A$ and $B$ are numerical constants to be determined in order to satisfy all our requirements. In the counterterm action, we have already fixed the coefficient of $\Phi^2$ so that the sum $I_{bulk} + I_{GHBY} + I_{ct}$ is finite and yields finite expectation values.\footnote{Note that this coefficient is different from that in the standard expression in eq.~\reef{boundary}.} However, these expressions do not satisfy the desired Ward identities.

Hence our goal will now be to determine $A$ and $B$ so that it is possible to get finite expectation values {\it and} satisfy both Ward identities. Note that we only have two free parameters to satisfy three (or four) different requirements. Getting finite expectation values for $\calo_{\Delta}$ and $T_{ab}$ will give $B=B(A)$. Satisfying the trace Ward identity will then fix $A$. If this procedure is consistent, and we will show it is, then we should be able to automatically satisfy the diffeomorphism Ward identity, that is independent from the other. Checking that the second Ward identity is satisfied then provides a highly nontrivial check of our results. At this point, let us add that in fact, the finite part of this resulting (finite) term \reef{wacko} yields 
\beq
I_{Legendre}= - \frac{\ve^2}{2\lp^{d-1}}\int d^dx\, (2\Delta-d) \, \phi_0 \, \phi_1 = - \int d^dx\, \lambda \, \vev{\calo_\Delta}\,.
\labell{power}
\eeq
Hence we have produce exactly the expression required for the Legendre transform in the boundary theory --- more details on this procedure are given in \cite{appear}.

So the first step is to compute the expectation values for the stress tensor and the operator and set $B$ as a function of $A$ by requiring them to be finite. Note that in this regime we will need to cancel terms proportional to $z_\epsilon^{-d+2\Delta}$, whereas in the usual holographic renormalization approach the counterterms are set to cancel the terms proportional to $z_\epsilon^{d-2\Delta}$. An important point in our calculation is that even though we are mostly treating $\phi_1$ and $\phi_0$ as independent in this perturbative analysis, in a more general setup $\phi_1$, proportional to the expectation value of $\calo_\Delta$, will be a function of $\phi_0$, the coupling. Then, if we want to compute the variation of some element ${\cal{X}}$ with respect to the source $\lambda$ we should consider both contributions,
\begin{equation}
\frac{\delta {\cal{X}}}{\delta \lambda } = \frac{1}{\ve} \left( \frac{\delta {\cal{X}}}{\delta \phi_0} +\frac{\delta {\cal{X}}}{\delta \phi_1} \frac{\delta \phi_1}{\delta \phi_0} \right)\,. \labell{phi0phi1}
\end{equation}
In the case of the usual holographic renormalization, the second term will be negligible, but will be important in this case. In fact, we will be assuming that $\phi_1$ is just proportional to $\phi_0$, \ie $\phi_1 = k_1 \phi_0$. This assumption is reasonable, at least at this order in the $\ve$ expansion, as any higher order terms will include additional powers of $\ve$. In any event, even though this will give contributions to the different parts of the action, the final result will not depend on this assumption.

Now we need to take variations with respect to the different terms in the regulated action. For that, it would be useful to consider,
\begin{eqnarray}
\vev{T^{ab}} & = & \lim_{z_\epsilon \to 0} \frac{2}{ \sqrt{-g^{(0)}}} \frac{\delta I_{reg}}{\delta g_{ab}^{(0)}} \nnn \\
& = & \lim_{z_\epsilon \to 0} \frac{2}{\sqrt{-g^{(0)}}} \left( \frac{\delta I_{reg}}{\delta \gamma_{cd}}\frac{\delta \gamma_{cd}}{\delta g_{ab}^{(0)}} + \frac{\delta I_{reg}}{\delta \Phi}\frac{\delta \Phi}{\delta g_{ab}^{(0)}} \right) \,, \labell{variation stress}\\ 
\vev{\calo_\Delta} & = & \lim_{z_\epsilon \to 0} \frac{1}{\sqrt{-g^{(0)}}} \frac{\delta I_{reg}}{\delta \lambda} \nnn \\
& = & \lim_{z_\epsilon \to 0} \frac{1}{\sqrt{-g^{(0)}}} \left(\frac{\delta I_{reg}}{\delta \gamma_{cd}} \frac{\delta \gamma_{cd}}{\delta \lambda} +  \frac{\delta I_{reg}}{\delta \Phi} \frac{\delta \Phi}{\delta \lambda} \right) \labell{variation op} \, .
\end{eqnarray}
To compute the variations we will be closely following Appendix C in \cite{anton}, fitting our formulas to the present simpler case where the scalar field does not have any dependence on the boundary spacetime coordinates. For instance,
\begin{eqnarray}
\frac{2 \lp^{d-1}}{\sqrt{-\gamma}}\,\frac{\delta\left( I_{bulk} + I_{GHBY}  \right)}{\delta
\gamma_{ab}} & = & \frac{z}{2} \left( \gamma^{ac}\gamma^{bd}\partial_{z}\gamma_{cd}
- \gamma^{ab}\gamma^{cd}\partial_{z}\gamma_{cd} \right)\Big{|}_{z=z_\epsilon} \,, \\
\frac{2 \lp^{d-1}}{\sqrt{-\gamma}}\,\frac{\delta\left( I_{bulk} + I_{GHBY}  \right)}{\delta
\Phi} & = & z \partial_{z}\Phi\Big{|}_{z=z_\epsilon}\,, \\
\frac{2 \lp^{d-1}}{\sqrt{-\gamma}}\,\frac{\delta I_{ct} }{\delta \gamma_{ab}}
 & = &  -\frac{1}{2}\gamma^{ab}\left( 2(d-1) + \frac{\Delta}{2}\Phi^2 \right)\Big{|}_{z=z_\epsilon}\,, \labell{eq app}\\
\frac{2 \lp^{d-1}}{\sqrt{-\gamma}}\,\frac{\delta I_{ct} }{\delta \Phi}
& = & - \Delta \, \Phi \Big{|}_{z=z_\epsilon}\,.
 \labell{eq:actionvar}
\end{eqnarray}
The variations of the Legendre action are also straightforward, \ie the variation with respect to the induced metric is analogue to eq. (\ref{eq app}), but with the term $B\, \Phi \, (\hat{n}\! \cdot\! \nabla) \, \Phi$. Then we need to recall eq. (\ref{phi0phi1}) when taking variations with respect to the source $\lambda$.

Now we can compute the expectation values of interest. The first requirement we want to impose is to have finite expectation values, so we will concentrate in the divergent terms, \ie again, in this regime, these are the ones proportional to $z_\epsilon^{-d+2\Delta}$. In the computation of the energy density, $T^{00}$, the divergent terms read as
\begin{eqnarray}
T^{00}_{div} = \frac{1}{ \lp^{d-1}} z_\epsilon^{-d+2\Delta} \phi_1^2 \ve ^2 ( B \, \Delta + A ) \,,
\end{eqnarray}
which gives 
\begin{eqnarray}
B= -\frac{A}{\Delta } \,. \labell{regulatingT}
\end{eqnarray}
Note that $B=0$ implies $A=0$, which, as mentioned, will give a finite $I_{reg}$. We might also consider choosing $A=-\Delta/2$ with which the term proportional to $\Phi^2$ in the Legendre action will cancel that in the counterterm action. This would produce a boundary term (with $B=1/2$) resembling that appearing in the original discussion of \cite{alter}. However, we will find that neither of these choices yields the desired Ward identities. It is important to mention that by satisfying eq. (\ref{regulatingT}), the expectation values of the spatial components of the stress tensor, as well as of  $\calo_\Delta$, also become finite.

Next step is to determine the  coefficient $A$ by requiring that the conformal Ward identity is satisfied. Given the relation, between $A$ and $B$, the finite expectation values for the operator and the trace of the stress energy tensor turn out to be,
\begin{eqnarray}
\vev{T^a{}_a} & = & - \frac{1}{2 \lp^{d-1}} \phi_0 \phi_1 \ve^2 \frac{(2 \Delta -d) (4A+\Delta^2)}{\Delta} \,, \\
\vev{\calo_\Delta} & = & - \frac{1}{2 \lp^{d-1}} \frac{(2A+\Delta) (2 \Delta -d) \phi_1 \ve}{\Delta} \,.
\end{eqnarray}
The conformal Ward identity reads
\begin{equation}
\vev{T^a{}_a} = (d-\Delta)\,\lambda\,  \vev{\calo_\Delta} \,,
\end{equation}
where $\lambda=\ve \phi_0$. Comparing these expressions, we conclude that
\begin{equation}
A = -\Delta  \qquad{\rm and\  hence}\quad B=1 \,.
\end{equation}
Substituting these values back into eq.~\reef{wacko}, we obtain the regulated action that appears in eqs.~(\ref{fullaction}--\ref{legendreaction}) in the main text. Of course, by construction, it is finite and satisfies the conformal Ward identity.

Still, the inclusion of the Legendre term is novel, so we would like to check whether with this action, we also satisfy the diffeomorphism Ward identity,
\begin{eqnarray}
\nabla^b\, T_{ab} = \vev{\calo_{\Delta}}\, \partial_a \lambda \,.
\end{eqnarray}
Of course this is trivial to satisfy in the case where neither the stress energy tensor nor the coupling have any space-time dependence. But we are looking for a nontrivial check. 
For that, we will assume now that all our coefficients have  some temporal dependence. Of course, in general this can set up a much more complicated problem such as the one considered in \cite{anton}. However, working at leading order in the perturbation parameter $\ve$, we will have some simplifications. In particular, we will assume that to leading order the expectation values won't be modified with the exception that $\phi_0,\phi_1$ and $m_{00}$ will be now functions of time. Then, we have,
\begin{eqnarray}
\vev{T_{00}(t)} = \frac{\ve^2}{2 \lp^{d-1}} \left( d \, m_{00}(t) - \frac{(d-\Delta)(2\Delta-d)}{d} \phi_0(t) \, \phi_1(t) \right)  \,,
\end{eqnarray}
and the Ward identity, as there is only time dependence, simplifies to
\begin{eqnarray}
\partial_t \vev{T_{00}(t)} = -\vev{\calo_{\Delta}(t)} \,\ve\,\partial_t \phi_0 (t) \,.\labell{bomb}
\end{eqnarray}
Now, it is obvious that the naive evaluation of the identity will not be satisfied, as the LHS will contain a term proportional to $m_{00}'(t)$ and the RHS will not.\footnote{In this section, $x'(t)$ will mean $\partial_t x(t)$. } However, we have an extra constraint coming from the Einstein's equations. By adding time dependence, there will be a new nontrivial $zt$ equation, $R_{zt}-\frac{1}{2} \partial_t \Phi \partial_z \Phi =0$, which determines $m_{00}'(t)$ in terms of $\phi_0'(t)$ and $\phi_1'(t)$. Computing that extra equation we obtain,
\begin{eqnarray}
m_{00}'(t) & = & -\frac{\Delta  (2 \Delta -d) }{d^2} \phi_1(t) \phi_0'(t) -\frac{(\Delta -d) (2\Delta-d)}{d^2} \phi_0(t) \phi_1'(t) \,.
\end{eqnarray}
Now it is now straightforward to show that the diffeomorphism Ward identity \reef{bomb} is perfectly satisfied if we impose this condition. 

The fact that we obtain a consistent Ward identity given our choice of $A$ and $B$ is highly nontrivial and provides an argument in favour of this alternate holographic renormalization procedure. After the whole process we obtained an action that provides finite expectation values that satisfy both Ward identities in the regime where $(d-2)/2<\Delta<d/2$. As far as our knowledge goes, this was not calculated before in the literature and it would be nice to have further checks of this proposal. 
For example, we can test that this generates the correct thermodynamics when we analyze the full black hole case \cite{appear}.
As for now, this shows that holography can deal with this kind of cases. Moreover, as detailed in the main text, this provides a confirmation that in this regime, the leading term in the entanglement entropy expansion for small spheres is not simply related to the stress energy tensor alone but rather involves $\delta\vev{\calo_\Delta}^2$.

Finally, let us comment on how the same action was derived in \cite{don2} without asking that the Ward identities be satisfied. Instead, the two requirements to fix $A$ and $B$ were finiteness of the on-shell action and stationarity of the action with either Dirichlet or Neumann boundary conditions. The scalar field part of the action is,
\begin{eqnarray}
I_{s} = -\frac{1}{2\lp^{d-1}} \int_M \sqrt{-G} \left( \frac{1}{2} (\nabla \Phi)^2 +\frac{1}{2} m^2 \Phi^2 \right)  - \frac{1}{2\lp^{d-1}} \int_{\partial M} \sqrt{\gamma} \left(\left(\frac{\Delta}{2}+A\right) \Phi^2 + B\, \Phi\ \hat{n}\! \cdot\! \nabla  \Phi \right) \nnn  \,.\\
\end{eqnarray}
Now, varying the bulk action by parts (with respect to $\Phi$) one gets equations on motion plus a boundary term. Then we impose the usual boundary behaviour to the scalar field, \ie $\Phi(z) = \ve \phi_0 z^{d-\Delta} + \ve \phi_1 z^d$.

By carefully analyzing the finite part of the action, we arrive to a variation that reads,
\begin{eqnarray}
\delta I_{s} = EOM - \frac{1}{2 \lp^{d-1}} \int_{\partial M} & & \left(  \left( 2A + B d \right) \phi_1\, \delta \phi_0  \right. \nnn \\
& & \quad\left. + \left( 2A+ B d-d+2\Delta   \right) \phi_0\, \delta \phi_1  \right)\,. \labell{stationary}
\end{eqnarray}
Now, imposing finiteness involves analyzing the divergent terms of the action and yields the same condition as found before in eq.~\reef{regulatingT}, \ie $B= -A/ \Delta $. 
However, as an additional constraint, we also want the action to be stationary under so-called Dirichlet (\ie $\phi_1$ fixed) or Neumann conditions (\ie $\phi_0$ fixed).\footnote{Recall that with $\Delta<d/2$, the (non)normalizable mode is associated with $\phi_0$ ($\phi_1$) in eq.~\reef{such2}.} For the Dirichlet boundary condition, we choose $A=0=B$ to remove the boundary term in the first line of eq.~\reef{stationary}. However, for the Neumann boundary condition, we need to cancel the second line which fixes $A=-\Delta$ and hence $B=1$ (just as found above by requiring the Ward identities to be satisfied). 

\section{$\vev{T_{00}}^2$ contribution for $d=2$ CFT}
\label{slider}
A simple calculation shows that a term proportional to the square of the energy density appears with a factor of $1/c$ in the entanglement entropy of a $d=2$ CFT. 
The entanglement entropy for an interval of length $2R$ in a thermal state is given by \cite{cardy0,korepin}
\beq
S(2R,T)=\frac{c}{3}\,\log\!\left[\,\frac{\sinh\left(2\pi  R\, T
\right)}{\pi T\delta}\,\right]\,,
 \labell{finiteT}
\eeq
where $c$, $T$ and $\delta$ are the central charge,\footnote{Here, we use the standard conventions for $d=2$ CFTs where
$\vev{T_{zz}(z) \,T_{ww}(w)}=\frac{c/2}{(z-w)^4}$. Note that the conventions in eq.~\reef{emt2p} yield $C_T=4c$.} temperature and short-distance cut-off, respectively.
Considering the shift from the vacuum entanglement entropy, we have
\beq
\delta S=S(2R,T)-S(2R,0)=\frac{c}{3}\log\left[\frac{\sinh(2\pi R\, T)}{2\pi R\, T}\right]\,,
\eeq
and in the regime $R\,T\ll1$, we may expand this expression to find
\beq
\delta S\simeq\frac{c}{18}(2\pi  R\, T)^2-\frac{c}{540} (2\pi R\, T)^4+\cdots \,. \labell{series}
\eeq
Further, the thermal energy density is given by
\beq
\vev{T_{00}} =\frac{\pi}{6}\, c\, T^2 \labell{edense}
\eeq
and hence eq.~\reef{series} becomes
\beq
\delta S=\frac{4\pi R^2}{3} \,\langle T_{00}\rangle-\frac{16 \pi^2}{15} \frac{R^4}c\, \langle T_{00}\rangle^2 +\cdots\,.
\eeq
The first term matches precisely the expected first law contribution given in eq.~\reef{IRS} for $d=2$. Two observations for the second order contribution are: first, the sign is negative ensuring that $\delta S\le \delta\vev{\cal H}$ and second, the coefficient appears with a factor of the inverse of the central charge. This factor is precisely analogous to those appearing in the second order contributions in eqs.~\reef{formula} and \reef{formula2}. We might also observe that this factor arises here because the expectation value of the energy density is itself proportional $c$, as shown in eq.~\reef{edense}. Similar formulae can be obtained holographically in any number of dimensions --- see section 3.2 of \cite{relative}. 

\section{Higher order corrections}
\label{higher}

In all of our computations, we have two different deformations: the first one is the deformation of the theory away from the conformal one; and the second is the excitation of the state away from the new vacuum. We are parameterizing these small deformations with our parameter $\ve$. In this Appendix, we will compute the next-to-leading order corrections, \ie the $\ve^3$ corrections, to the entanglement entropy. 
The procedure is analogous to the calculations in section \ref{general} but with a series of modifications. First, we need to include $\kappa \neq 0$ in the scalar potential \reef{potential}. Without this cubic term,  we would not find an extra contribution to $\delta S$ at order $\ve^3$. Then, since the scalar field equation \reef{KGeq2} is now nonlinear, we need to include cross-terms between the $\phi_0$ and the $\phi_1$ series in the scalar field expansion. 

Note that while the following analysis is valid for generic values of $\Delta$, at this order, there will be certain special values (\eg $\Delta=2d/3$) that need to be analyzed separately because of the appearance of extra logarithms, as appeared for $\Delta=d/2$ in section \ref{deltaover2}. We do not contemplate those cases in the rest of this Appendix. 

At order $\ve^2$, the ansatz for the scalar field will have the most general form 
\begin{eqnarray}
\Phi(z) & = & \phi_0 \ve  z^{d-\Delta } + \phi_1 \ve  z^{\Delta } + f_1\, \phi_0^2 \ve^2  z^{2(d-\Delta) }+ f_2\, \phi_1^2 \ve^2  z^{2\Delta } + f_3 \,\phi_0 \phi_1 \ve^2 z^d   \,.
\end{eqnarray}
As before, $\phi_0$ and $\phi_1$ will be free parameters related to the coupling and the expectation value of $\cO_\Delta$ and the $f_i$'s are constants to be determined by the equations of motion. The ansatz for the metric is completely analogous, \ie we write all the possible $\ve^3$ terms possible. We report here just the answer obtained after solving the equations to order $\ve^3$ (and linear order in $\kappa$). The scalar field solution can be written as
\begin{eqnarray}
\Phi ( z)  & = & \phi_0 \, \ve \, z^{d-\Delta }+\phi_1 \, \ve \,  z^{\Delta } + \\
& & \quad+  \kappa \, \ve^2 \left(\frac{\phi_0^2 \, z^{2 (d-\Delta )}}{2 (2d-3 \Delta ) (d-\Delta )}- \frac{\phi_1^2 \, z^{2\Delta }}{2 \Delta  (d-3 \Delta  )}+\frac{\phi_0 \phi_1 \, z^d}{\Delta  (d-\Delta )} \right) \,. \nnn
\end{eqnarray}

For simplicity, we limited the present calculations to a state described by a static and isotropic bulk geometry, \ie $g_{0i}(x)=0$ and $g_{ij}(z)\propto\delta_{ij}$ in eq.~\reef{fgump}, but our results for the third order corrections are still general. At second order, this choice corresponds to restricting the coefficients in eq.~\reef{juk} to have a simple diagonal form and in the end, it only affects the solution for $m_{ab}$. The third order corrections for the metric have a simple form with: $g_{ab}(z) = g_{ab}^{(\ve^2)} + \ve^3\,\eta_{ab}\, g_3 (z)$ where\footnote{We thank G. Sarosi and T. Ugajin for pointing out a mistake in this formula in a previous version of this draft.}
\begin{eqnarray}
& & g_3(z) = \frac{2 \, \kappa }{(d-1)} \left( \frac{\phi_1^3\, z^{3 \Delta }}{9 \Delta  (d-3 \Delta )} - \frac{\phi_0^3 \, z^{3 (d-\Delta )}}{9 (d-\Delta ) (2 d-3 \Delta )} \right. \\
& & \qquad\qquad\qquad\left.-\frac{\Delta \, \phi_0 \, \phi_1^2 \, z^{d+\Delta }}{\left(d^2-\Delta ^2\right) (d-3 \Delta )} -\frac{(d-\Delta ) \, \phi_0^2 \, \phi_1 z^{2 d-\Delta }}{\Delta  (2 d-\Delta ) (2 d-3 \Delta )} \right) \,. \nnn
\end{eqnarray}
Of course, the functions at order $\ve^2$ are precisely the same as found in section \ref{general}. Hence
the solution goes back to the previous one if we set $\kappa=0$ and restrict to terms of order up to $\ve^2$.

With the perturbation of the metric in hand, we can now proceed to compute the entanglement entropy. However, a few comments are in order before that though. First, in the holographic computation of the entanglement entropy for perturbed states, we have two different contributions: one related to the change in the background metric and the other related to the change in the position of minimal surface. The leading metric corrections appear at order $\ve^2$. This means that corrections in the position of the extremal surface (the semi-sphere for CFT's vacuum) will appear at this same order but the leading corrections which this shift in the position makes to $\delta S$ appear at order $\ve^4$. Of course, this argument tells us that we can neglect that contribution for the current computation, which only considers corrections up to order $\ve^3$.

Now let us write $\delta S$ in the form
\begin{equation}
\delta S = \frac{2\pi \Omega_{d-2}}{\lp^{d-1}} \int_0^R \left( \ve^2\,\Delta s_2(z)  + \ve^3\,\Delta s_3(z)  \right) dz \,,
\end{equation}
where we already integrated the angular coordinates. Of course, $\Delta s_2 (z)$ is the leading order contribution that we already computed in eq. (\ref{minarea}). After some algebra, $\Delta s_3(z)$ 
turns out to be
\begin{eqnarray}
\Delta s_3(z) & = & \kappa \,  z \frac{\left(R^2-z^2\right)^{\frac{d-3}{2}}}{R(d-1)} \, \left((d-2) R^2+z^2\right)  \\
& & \times \left( \frac{\delta \phi_1^3\, z^{3 \Delta -d}}{9 \Delta  (d-3 \Delta )} -\frac{\Delta \, \phi_0 \, \delta\phi_1^2 \, z^{\Delta }}{\left(d^2-\Delta ^2\right) (d-3 \Delta )} -\frac{(d-\Delta ) \, \phi_0^2 \, \delta\phi_1 z^{d-\Delta }}{\Delta  (2 d-\Delta ) (2 d-3 \Delta )} \right) \,. \nnn
\end{eqnarray}
Again, the integrand is everywhere finite so there is no need to introduce a cut-off. After integration, the entanglement entropy receives three types of $\ve^3$-contributions,
\begin{eqnarray}
\delta S^{\ve^3} =  \delta S_1^{\ve^3} +\delta S_2^{\ve^3} +\delta S_3^{\ve^3} \,,
\end{eqnarray} 
where
\begin{eqnarray}
\delta S_1^{\ve^3} & = & - \frac{\pi \Omega_{d-2}}{\lp^{d-1}} \delta\phi_1^3 \, \kappa \, R^{3 \Delta } \frac{\Gamma \left(\frac{d-1}{2}\right) \Gamma \left(\frac{3 \Delta -d}{2}\right)}{12 \, \Gamma \left(\frac{3 \Delta +3}{2}\right)} \,,\nonumber\\
\delta S_2^{\ve^3} & = & -\frac{\pi \Omega_{d-2}}{\lp^{d-1}} \phi_0 \, \delta\phi_1^2 \, \kappa \, R^{d+\Delta}   \frac{\Delta  \Gamma \left(\frac{d-1}{2}\right) \Gamma \left(\frac{\Delta +2}{2}\right)}{2 (d-3 \Delta ) (d-\Delta ) \Gamma \left(\frac{d+\Delta +3}{2}\right)} \,,\label{contin}\\
\delta S_3^{\ve^3} & = & - \frac{\pi \Omega_{d-2}}{\lp^{d-1}} \phi_0^2 \, \delta\phi_1 \, \kappa \, R^{2 d-\Delta } \frac{(d-\Delta ) \Gamma \left(\frac{d-1}{2}\right) \Gamma \left(\frac{d-\Delta +2}{2}\right)}{2 (2 d-3 \Delta ) \Delta \Gamma \left(\frac{2d-\Delta+3}{2}\right)} \,.\nonumber
\end{eqnarray}
Each of the three contributions are proportional to the cubic coupling in the scalar potential $\kappa$. Hence we see that it was important to add that term in order to get a nonzero result at this order in the $\ve$ expansion. 

Next we would like to write $\delta S$ in terms of field theory quantities. For that, we should carry out the holographic renormalization procedure as in section \ref{holren}. As we introduced an extra cubic term in the scalar potential, we should expect that new divergences will arise and that we need to modify our counterterm action in order to remove them. In fact, one can show that a finite action is produced with
\begin{eqnarray}
I_{ct} & = & - \frac{1}{2 \lp^{d-1}} \int d^d x \left. \sqrt{-\gamma} \left( 2(d-1) + \frac{d-\Delta}{2} \Phi^2 + \frac{\kappa}{6(2d-3\Delta)} \Phi^3  \right) \right\rvert_{z=z_\epsilon} \, .\end{eqnarray}
Next, we need to compute the expectation value for the stress tensor and the scalar operator. However, this computation does not produce any extra contributions to either the expectation value of the stress tensor or of the operator. So the expectation values are unchanged at order $\ve^3$. Using eq.~(\ref{vevO}), we can rewrite the entanglement entropy contributions \reef{contin} as functions of the field theory coupling and expectation value. In doing so, we find
\begin{eqnarray}
\delta S_1^{\ve^3} & = & - \frac{1}{\pi \, \Omega_{d-2}} \frac{\delta \vev{\calo_\Delta}^3}{C_T^{\,2}} \, \kappa \, R^{3 \Delta } \frac{2^{2 d-1} d^2 (d+1)^2 \Gamma \left(\frac{d-1}{2}\right) \Gamma \left(\frac{3 \Delta - d}{2}\right)}{3 (2 \Delta -d)^3 \Gamma \left(\frac{3 \Delta +3}{2}\right)} \,,\\
\delta S_2^{\ve^3} & = & - \lambda \, \frac{\delta \vev{\calo_\Delta}^2}{C_T} \, \kappa \, R^{d+\Delta}   \frac{2^d d (d+1) \Delta  \Gamma \left(\frac{d-1}{2}\right) \Gamma \left(\frac{\Delta +2}{2}\right)}{(d-3 \Delta ) (2 \Delta-d )^2 (d-\Delta ) \Gamma \left(\frac{d+\Delta +3}{2} \right)}\,,\\
\delta S_3^{\ve^3} & = & -\pi \, \Omega_{d-2} \, \lambda^2 \, \delta\vev{\calo_\Delta} \, \kappa \, R^{2 d-\Delta } \frac{(d-\Delta ) \Gamma \left(\frac{d-1}{2}\right) \Gamma \left(\frac{d-\Delta +2}{2}\right)}{\Delta  (2 d-3 \Delta ) (2 \Delta -d) \Gamma \left(\frac{2d - \Delta +3 }{2}\right)} \,. 
\end{eqnarray}

So, in all, we have three different possible contributions to order $\ve^3$, each one proportional to a different power of the expectation value for the operator. Note that the last two terms will give higher order contributions in the $R$ expansion when $\Delta<d$, but depending on the space-time dimension, the first term can also be important when $(d-2)/2 < \Delta < d/2$. Of course, an alternative renormalization procedure would be needed in that case to make sure that the term remains unchanged in that regime. But assuming this is true, then that term would dominate over $R^d$ contributions in the small $R$ expansion. We can write $\Delta$ as $\Delta=d/2-\nu$, with $0<\nu<1$. Then the first term would scale as $R$ to the power $3d/2 - 3\nu$, that consequently gives that a power of $R$  smaller than $d$ if $\nu>d/6$. Of course for this to happen, $d<6$ (because $\nu<1$). Even in that case, there will be a term proportional to $\ve^2$ that scales as $R^{2\Delta}$, that would be the leading contribution to the entanglement entropy.

\end{document}